\documentclass[prd,twocolumn,floatfix,preprintnumbers,altaffilletter,superscriptaddress,nofootinbib]{revtex4-2}
\usepackage{graphicx,url,amssymb,amsmath,rotating,color,units,wasysym,epsfig,multirow,epstopdf,subcaption}
\usepackage[colorlinks=true,citecolor=blue,urlcolor=blue]{hyperref}

\usepackage{soul,xcolor}
\usepackage{aas_macros}

\usepackage{algpseudocode}
\usepackage[ruled]{algorithm}
\usepackage{booktabs}

\def\code#1{\texttt{#1}}

\graphicspath{{.}{./figures/}{./figures/toy_model/}{./figures/initialization_comparison/}}

\begin{document}

\title{Accelerating Tests of General Relativity with Gravitational-Wave Signals using Hybrid Sampling}

\author{Noah E.~Wolfe}
\email{noah.wolfe@ligo.org}
\affiliation{Department of Physics, North Carolina State University, Raleigh NC 27695, USA}
\affiliation{Department of Mathematics, North Carolina State University, Raleigh NC 27695, USA}

\author{Colm Talbot}
\email{colm.talbot@ligo.org}
\affiliation{LIGO Laboratory, Massachusetts Institute of Technology, 185 Albany St, Cambridge, MA 02139, USA}
\affiliation{Department of Physics and Kavli Institute for Astrophysics and Space Research, Massachusetts Institute of Technology, 77 Massachusetts Ave, Cambridge, MA 02139, USA}

\author{Jacob Golomb}
\affiliation{LIGO Laboratory, California Institute of Technology, Pasadena, CA 91125, USA}

\date{\today}

\begin{abstract}

The Advanced LIGO/Virgo interferometers have observed $\sim 100$ gravitational-wave transients enabling new questions to be answered about relativity, astrophysics, and cosmology.
However, many of our current procedures for computing these constraints will not scale well with the increased size of future transient catalogs.
We introduce a novel hybrid sampling method in order to more efficiently perform parameterized tests of general relativity with gravitational-wave signals.
Applying our method to the binary black hole merger GW150914 and simulated signals we find that our method is approximately an order of magnitude more efficient than the current method with conservative settings for our hybrid analysis.
While we have focused on the specific problem of measuring potential deviations from relativity, our method is of much wider applicability to any problem that can be decomposed into a simple and more complex model(s).

\end{abstract}

\maketitle

\section{Introduction}

General relativity (GR) is currently our most successful theory of gravity. Previous observations of sources within our solar system, including the Gravity Probe B experiment and time-delay measurements with the Cassini space probe, have placed constraints on deviations from general relativity in the non-dynamical, weak-field regime~\citep{gr-tests_will_2014}.
These tests have been replicated and expanded with radio observations of pulsars, which probe similarly slow-motion, but strong-field gravitational physics \citep{pulsar-timing-tgr_ingrid_2003, pulsar-tgr_wex_2014}, through measurements of the orbital decay rate of the first discovered binary pulsar system~\cite{hulse-taylor-tgr_damour_1974} to modern constraints on dipolar gravitational-wave emission constructed with multiple such systems \cite{binary-pulsar-tgr_zhao_2022}.
Simultaneously, probes of large-scale cosmological structure including weak gravitational lensing and the Cosmic Microwave Background have provided complimentary weak-field tests of general relativity across cosmic epochs and length scales \cite{cosmo-tgr-review_ferreira_2019}.
Over the past decade new observations have unlocked the strong-field regime for tests of general relativity, including measurements of two supermassive black hole shadows \citep{EHT_M87_gr-test, EHT_SgrA_metric, EHT_SgrA_shadow} and gravitational waves from stellar-mass compact object mergers observed by Advanced LIGO~\cite{LIGO} and Advanced Virgo~\cite{Virgo}, using both single observations \cite{single-BBH_gr-tests_LIGO_2016, single-NSNS_gr-tests_LIGO_2019} and the burgeoning population of gravitational wave transients \cite{O1_gr-tests_LIGO_2016, GWTC1_gr-tests_LIGO_2019, GWTC2_gr-tests_LIGO_2020, GWTC3_gr-tests_LIGO_2021}.
To date, none of these experiments have found significant disagreement with the predictions of general relativity.

However, alternative theories of gravity that could emerge in the strong-field regime may be relevant to constructing unified field theories or the understanding of unexplained phenomena like dark energy (e.g. in scalar-tensor theories, among others \cite{scalar-tensor-review}).
Modern developments in theoretical physics have generated testable predictions of modifications to gravitational-wave emission from compact binary coalescence under alternative formulations of gravity, both analytically \citep{ppE_Yunes_2009, analytical-scalar-tensor_lang_2014, analytical-scalar-tensor_lang_2015, analytical-scalar-tensor-gw_sennett_2016, ppe-waveforms_tahura_2018, ppE_bonilla_2022, edgb-analytical_shiralilou_2022} and numerically~\citep{bns-scalar-tensor_barausse_2013, bns-scalar-tensor_shibata_2014, dCS_nr_okounkova_2017, dCS-nr-gw150914_okounkova_2020, edgb-nr-gw150914_okounkova_2020, eft-gravity_cayuso_2020, edgb-nr_east_2021, edgb-nr-bbh_east_2021}, further enabling tests of general relativity in the most extreme gravitational environments yet accessible to us.

The number of observed mergers will only continue to grow, and our gravitational-wave detectors will continue to become more sensitive, further enhancing our resolution on potential deviations from general relativity.
However, this also necessitates that our statistical and computational techniques improve to support larger and more complex analyses.
Since the first observation of gravitational waves from a compact binary merger, the ``TIGER''(Test Infrastructure for GEneral Relativity) formalism and related methods have been some of the flagship analyses performed by the LIGO-Virgo-KAGRA scientific collaborations~\citep{Agathos2014,O1_gr-tests_LIGO_2016,single-NSNS_gr-tests_LIGO_2019,GWTC1_gr-tests_LIGO_2019, GWTC2_gr-tests_LIGO_2020, GWTC3_gr-tests_LIGO_2021,Mehta2022,meidam2018}.
These methods require performing many independent, but largely identical analyses as for each potential deviation from relativity, the parameters describing the GR signal must be inferred from scratch.
The goal of this work is to improve this analysis procedure with a new method for parameter estimation: hybrid sampling.
Here, our hybrid approach uses an analysis assuming that general relativity is correct to initialize the inference of deviations from general relativity.
This method is more computationally efficient, allowing us to scale our analysis as the population of observed mergers grows, and further constrain deviations in gravitational-wave signals predicted by general relativity.

The remainder of the paper is structured as follows.
In Section \ref{sec:stat_methods}, we provide relevant background and introduce our hybrid sampling method.
After this, we provide a demonstration of our method on a simple toy model in Section~\ref{sec:demonstration}.
We then describe our model for observed gravitational waves according to general relativity and the parameterized deviations we consider in Section~\ref{sec:astro_methods}.
In Section~\ref{sec:hybrid-sampling_gws}, we apply our hybrid sampling method to simulated and real gravitational-wave signals.
Specifically, we demonstrate that our method returns equivalent results to the previous method at a fraction of the computational cost and introduce an extension to the previous method.
Finally, we provide closing thoughts in Section~\ref{sec:conclusions}.

\section{Methods} \label{sec:stat_methods}

\subsection{Bayesian Inference for Gravitational-Wave Transients}

We begin with a brief review of Bayesian inference in the context of gravitational-wave astronomy.
In Bayesian inference, we wish to infer a set of parameters $\theta$ of a model $M$ given some data $d$; formally, we want to construct the posterior distribution $p(\theta | d, M)$.
For example, in this work, we will have a set of parameters that include properties of binary black hole systems (e.g.~mass and spin), with additional parameters to denote deviations from the predictions of general relativity that we wish to infer from observations of gravitational-wave transients.
For additional details, see, e.g.~\cite{bayesian_inference_gws_thrane_2019}.

Bayesian inference allows us to construct the posterior distribution via Bayes' theorem,
\begin{equation} \label{eq:bayes}
    p(\theta | d, M) = \frac{ \mathcal{L}(d|\theta, M) \pi(\theta | M) }{ \mathcal{Z}(d | M) }
\end{equation}
where $\mathcal{L}(d|\theta, M)$ is the likelihood of observing the data given parameter values, and $\pi(\theta | M)$ is the prior distribution, which encodes our assumptions about the Universe before considering the data.
The normalization factor $\mathcal{Z}(d | M)$ is known as the evidence and is the probability of observing the data given the parametric model we choose
\begin{equation} \label{eq:evidence}
    \mathcal{Z}(d | M) \equiv \int d\theta \mathcal{L}(d | \theta, M) \pi(\theta | M).
\end{equation}
We may suppress the model $M$ in subsequent expressions, however, everything is conditioned on a model.

When analyzing gravitational-wave transients we assume that the noise in each of our interferometers is a stationary Gaussian process described by a power spectral density $S$ in the frequency domain.
Additionally, our analysis is triggered by matched filter search pipelines that tell us a coherent non-Gaussian transient that is most likely an astrophysical signal is present in the data.
To model this, we use the Whittle likelihood approximation~\cite{Romano2017} for the residual noise after subtracting the response of each detector to our template $h$ for the astrophysical signal
\begin{equation}
    {\cal L}(d | \theta) = \prod_{i,j} \frac{1}{2\pi S_{ij}} \exp\left( -\frac{4}{T}\frac{|d_{ij} - h(\theta)_{ij}|^2}{S_{ij}} \right).
\end{equation}
Here, the products run over the interferometers in the network, and frequencies for the data in each interferometer are (generally) assumed to be uncorrelated.
We note that our parameters only describe the astrophysical template and the response of the detector, however, it is also possible to construct parameterized models for the power spectrum \citep{Littenberg2015}.
The quantity $T$ is the duration of data being analyzed and is the inverse of the frequency resolution.

We observe that $p(\theta | d)$ provides a distribution on the entire (multi-dimensional) set of parameters $\theta$.
To extract information on specific parameters of interest $\theta_i$, we must ``marginalize", i.e.~integrate, over the rest of the parameters:
\begin{equation}
    p(\theta_i | d) = \int \left( \prod_{k \neq i} d \theta_{k} \right) p(\theta | d).
\end{equation}
This integration may be difficult to compute through standard numerical methods, especially in a high-dimensional parameter space.
One common method to approximate this integral is to utilize a Markov chain Monte Carlo (MCMC) \citep{metropolis1953, hastings1970}, wherein a ``walker" explores the parameter space of $\theta$ under rules such that, given enough iterations, the combined steps along its path form a representative sample of the posterior distribution.
Another, more recent method is nested sampling \citep{skilling_2004, skilling_2006}, which instead focuses on estimating the evidence $\mathcal{Z}$, from which the posterior distribution can then be calculated.
In this work, we will utilize both of these approaches and, in turn, detail specific implementations of these methods in the following subsections.

\subsection{Nested Sampling} \label{sec:dynesty}

Nested sampling, as developed in \citep{skilling_2004, skilling_2006}, is an algorithm to estimate the evidence $\mathcal{Z}$ and posterior probability density by climbing up discrete contours on the likelihood surface and has been widely adopted in astrophysics including gravitational-wave astronomy~\cite{Veitch2015, Ashton2019, speagle_2020_dynesty, RomeroShaw2020}.
We direct interested readers to~\cite{Ashton2022} for a recent review.
The core insight of nested sampling is that the high dimensional integral to compute the evidence ${\cal Z}$ can be approximated as a one-dimensional integral over a quantity known as the ``prior mass'' $X$.
The prior mass corresponding to a likelihood value $\lambda$ is the fraction of the volume that has a likelihood greater than $\lambda$
\begin{equation}
    X(\lambda) = \int_{\mathcal{L}(\theta) > \lambda} \pi(\theta) d\theta.
\end{equation}

If the mapping from $\theta \rightarrow X$ can be found, the evidence (Eq.~\ref{eq:evidence}) can be rewritten as
\begin{equation}
    \mathcal{Z} = \int_0^{1} \mathcal{L}(X) dX.
\end{equation}
The nested sampling algorithm constructs this mapping numerically by gradually climbing the likelihood surface and we approximate $\mathcal{Z}$ as a weighted sum of values $\mathcal{L}(X)$, e.g.,
\begin{equation} \label{eq:nested_z_approx}
    \mathcal{Z} \approx \sum_{i=1}^{N} w_i \mathcal{L}_i
\end{equation}
for some number of samples $N$, with the prior volume associated with each likelihood iso-contour $w_i$.
For a full derivation of the functional form of $w_{i}$, see, e.g.,~\cite{skilling_2006}.
For this work, we use the implementation of nested sampling in \code{dynesty}~\cite{speagle_2020_dynesty}.

Another widely used feature of nested sampling is that the elements in the sum in Eq.~\ref{eq:nested_z_approx} are the posterior weights associated with nested sampling.
We can therefore generate samples from the posterior distribution by weighting the nested samples according to a normalized version of that quantity
\begin{equation}
    \label{eq:posterior_weights}
    p_i = \frac{\mathcal{L}_i w_i}{\mathcal{Z}}.
\end{equation}
We note that after a sufficient number of iterations, nested sampling no longer produces additional posterior samples.
This is because the algorithm continually climbs the likelihood surface and eventually the reduction in prior volume overcomes the increase in the likelihood and the posterior weights begin to decrease.
This means that the number of posterior samples generated by a nested sampling analysis is completely determined by the shape of the likelihood surface.

\begin{figure}
    \centering
    \includegraphics[width=\linewidth]{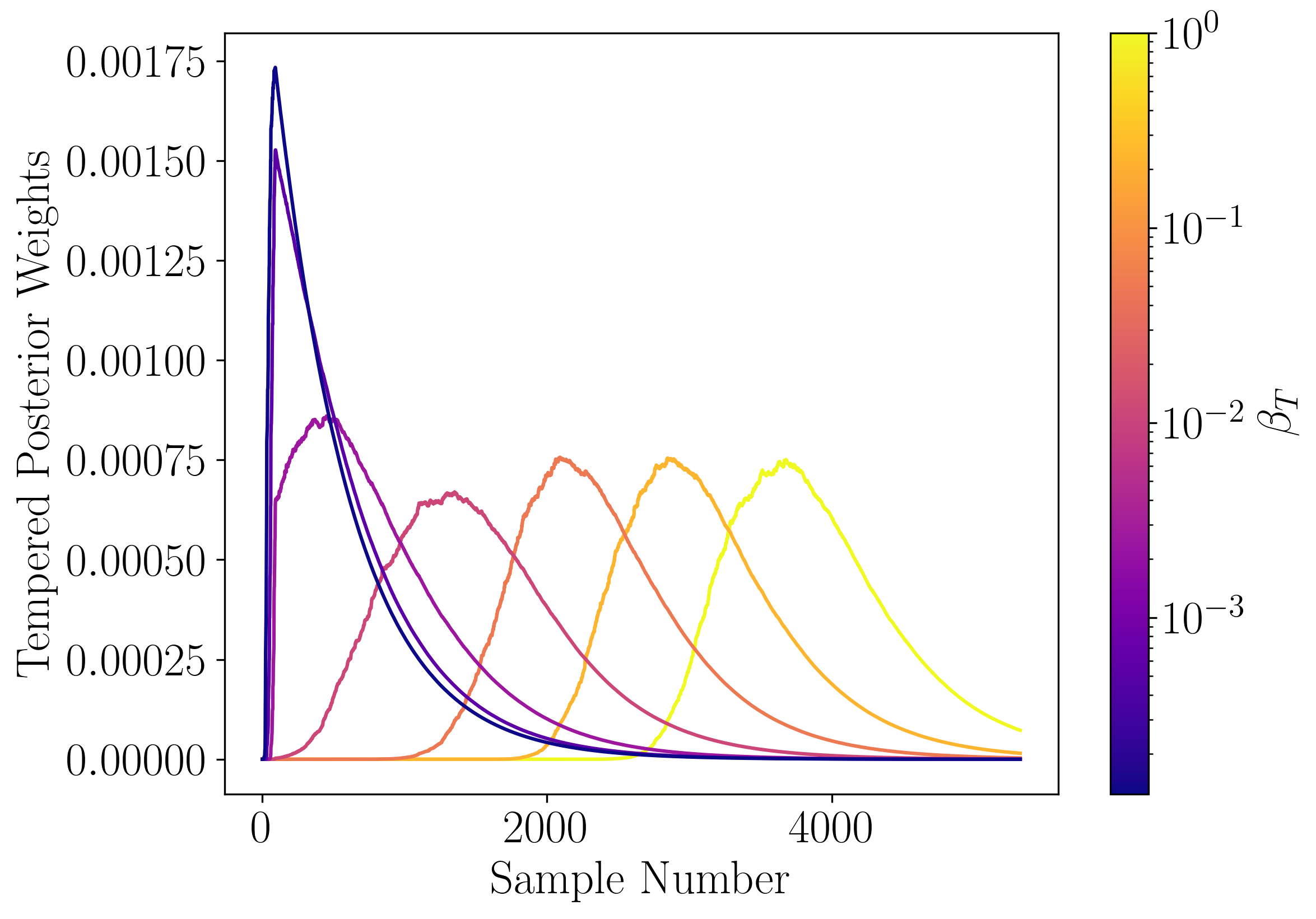}
    \caption{
        Posterior weights $p_i$ generated by \code{dynesty}, tempered according to the default method described in \cite{Vousden_2015_ptemcee}.
        As $\beta_T \rightarrow 1$, we recover the original distribution of posterior weights.
        As $\beta_T \rightarrow 0$ we recover an exponential distribution, the result of choosing nested samples according to $X_i = \exp(-i / N)$.
    }
    \label{fig:temperature_ladder}
\end{figure}

We note that the value of the ${\cal L}_{i}$ are irrelevant to the nested sampling algorithm, and that only their order matters~\footnote{Strictly speaking, this is only true if one has an infinite chain of nested samples.}.
Therefore, we are free to perform any monotonic operation on the likelihood and can then trivially recompute the evidence and generate samples from the posterior distribution.
In this work, we will focus on a specific family of operations that change the effective inverse ``temperature'' $\beta_{T}$ of the posterior distribution~\footnote{The temperature, in this case, is defined in analogy with statistical physics which historically shares strong links with Monte Carlo analyses.}
\begin{align}
    {\cal{L}} &\rightarrow {\cal L}^{\beta_{T}} \\
    p_{i,\beta_{T}} &= \frac{\mathcal{L}^{\beta_{T}}_i w_i}{\mathcal{Z}_{\beta_{T}}}. \label{eq:tempered-weights}
\end{align}
This ``athermal'' property of nested sampling has been known since the first introduction of the algorithm but has not been widely utilized.

In Figure~\ref{fig:temperature_ladder}, we show the posterior weight $p_{i,\beta_{T}}$ as a function of iteration for various temperatures for a simple model described in Section~\ref{sec:demonstration}.
We note that for the $\beta_{T} = 1$ case, we recover the usual posterior weights.
As we increase the temperature (decreasing $\beta_{T}$) the posterior weights peak earlier in the nested sampling chain.

\subsection{Parallel-Tempered Markov chain Monte Carlo} \label{sec:ptemcee}
In contrast with nested sampling, MCMC methods directly explore the posterior and can be run as long as necessary, continually generating additional samples from the posterior distribution.
Ensemble MCMC methods build upon existing MCMC methods by replacing a single walker, as used in traditional approaches~\citep{metropolis1953, hastings1970}, with an ensemble of walkers that explore the parameter space in parallel, e.g.,~\cite{goodman2010_ensemble}.
A key feature of an ensemble approach is that we can reduce the number of iterations we need to evolve the MCMC to accurately resolve the posterior distribution, as ensembles of walkers have a far shorter auto-correlation length, measuring the correlation between sampling steps, than single walkers \citep{goodman2010_ensemble, emcee_Foreman-Mackey_2013}.
Additionally, at any step, the state of our ensemble is a representative estimate of the posterior distribution.
For a recent review, we direct the readers to~\cite{Hogg2018}.
In further contrast with nested sampling techniques, we can choose the initial distribution of points in our ensemble.
Common initialization schemes include drawing random samples from the prior distribution and initializing samples around a maximum likelihood estimate, however, the optimal initialization is a realization of the target distribution.

Further, ensemble MCMC methods can be parallel-tempered to explore the posterior distribution at arbitrary temperatures \citep{swendsen1986}.
A parallel-tempered ensemble MCMC method then uses many walkers, in parallel, each exploring a tempered posterior surface
\begin{equation} \label{eq:continuous-tempered-posterior}
    p_{\beta_{T}}(\theta | d) = \frac{\mathcal{L}^{\beta_{T}}(d | \theta) \pi(\theta)}{{\cal Z}_{\beta_{T}}(d)}.
\end{equation}
We note that this is a continuous version of Eq.~\ref{eq:tempered-weights}.
For higher temperatures (smaller $\beta_{T}$), the ensemble can more easily explore the full prior volume, and by allowing walkers to jump between different temperature ensembles the convergence time of the $\beta_{T}=1$ ensemble is greatly reduced.
In this work, we use the \code{ptemcee} implementation of parllel-tempered ensemble MCMC \cite{emcee_Foreman-Mackey_2013, Vousden_2015_ptemcee}. 

While MCMC methods may not always be as computationally efficient as nested sampling methods, once they have reached a stationary state, they have a relatively high computational efficacy; i.e.~we can continually ask our MCMC sampler for more samples, increasing our resolution of the posterior as much as we desire.
Therefore, if we can initialize our MCMC ensembles to closely approximate the target distributions, we can achieve very large efficiencies.

\subsection{Hybrid Sampling} \label{sec:hybrid-sampling}

We propose a hybrid sampling scheme to explore high-dimensional, degenerate parameter spaces that uses the exponential compression of the prior mass provided by nested sampling to seed a set of parallel-tempered MCMC ensembles.
In particular, we seed each tempered ensemble by rejection sampling the nested samples, weighted by the tempered posterior weights defined by Eq.~\ref{eq:tempered-weights}, which approximates the tempered posterior distribution defined by Eq.~\ref{eq:continuous-tempered-posterior}.
In this work, we temper our MCMC ensembles over the default temperature ladder used by \code{ptemcee}, as detailed in Section 2.1 of \cite{Vousden_2015_ptemcee}.
In the case that the nested sampling is unbiased, this procedure initializes each ensemble from a realization of their target distribution.
This is the optimal seeding for the MCMC ensembles.
We can then run the tempered ensembles to generate an arbitrary number of samples from the target distribution (in constrast to nested sampling which can only generate a fixed number of samples, although dynamic nested sampling has been provides another solution to this problem~\cite{dynamic-nested-sampling}).

Our method can also be applied to more complex cases where the MCMC evolution explores an extended parameter space compared to the nested sampling analysis.
We define two models $M_{1}$ and $M_{2}$ described by parameter sets $\theta_{1} \subseteq \theta_{2}$ with likelihoods ${\cal L}_{1}$ and ${\cal L}_{2}$.
We denote the extension parameters as $\bar{\theta}$; i.e. $\theta_2$ is the union of $\theta_1$ and $\bar{\theta}$.
We first perform a nested sampling analysis of the data under model $M_1$, generating the posterior distribution $p(\theta_1 | d, M_1)$.
Since $M_2$ contains $M_1$, there must exist some value of $\bar{\theta}$ for which $M_2$ reduces to $M_1$, which we call $\bar{\theta}'$.
Therefore, $p(\theta_1 |d, M_1) = p(\theta_2 | d, M_2, \bar{\theta}')$, and we can consider other realizations of $p(\theta_2 | d, M_2)$ as an extension of the distribution for which $\bar{\theta} = \bar{\theta}'$.
So, the posterior distribution for $\theta_1$ that we achieve via nested sampling provides an efficient starting state for MCMC ensembles sampling in $\theta_2$ under $M_2$.
For the parameters of $\theta_2$ included in $\theta_1$, we seed each tempered ensemble as in the case where $M_1 = M_2$.
In the remaining parameters $\bar{\theta}$, we initialize our chains from narrow distributions centered around $\bar{\theta}'$.

In the rest of this work, we will take $M_1$ to be a model of the gravitational-wave signal from a binary black hole merger in accordance with general relativity, and $M_2$ to be a model of the same phenomenon that allows for deviations from GR. 
Then, $\bar{\theta}$ are parameters of these deviations, and $\bar{\theta}'$ is zero in each deviation parameter.

\section{Hybrid Sampling with a Generalized Gaussian Distribution}\label{sec:demonstration}

As a demonstration of our hybrid sampling framework, we use a toy model where our reduced model in the first step is the standard Gaussian distribution characterized by mean $\mu$ and standard deviation $\sigma$ and the complex model in the second step is a generalized Gaussian distribution characterized by mean $\mu$, scale $\alpha$, which can take alternative shapes parameterized by $\beta$.
For comparison, the probability density function of the standard Gaussian is
\begin{equation}
    P(x | \mu, \sigma) = \frac{1}{\sqrt{\pi \alpha^2}} e^{-(x - \mu)^2 / \alpha^2}
\end{equation}
while that of the generalized Gaussian we employ is
\begin{equation}
    P(x | \mu, \alpha, \gamma) = \frac{\beta}{2 \alpha \Gamma(1 / \gamma)} e^{-(|x - \mu|/\alpha)^{\gamma}}
\end{equation}
where $\Gamma$ is the Gamma function.
For consistency with the generalized model, we parameterize our Gaussian distribution with the parameter $\alpha = \sqrt{2} \sigma$.
When the shape $\gamma = 1$, we recover the Laplace distribution, while as $\gamma \rightarrow \infty$, we recover a tophat distribution on $(\mu - \alpha, \mu + \alpha)$.
When $\gamma = 2$, we recover the standard Gaussian distribution.

\begin{figure}
    \centering
    \includegraphics[width=\linewidth]{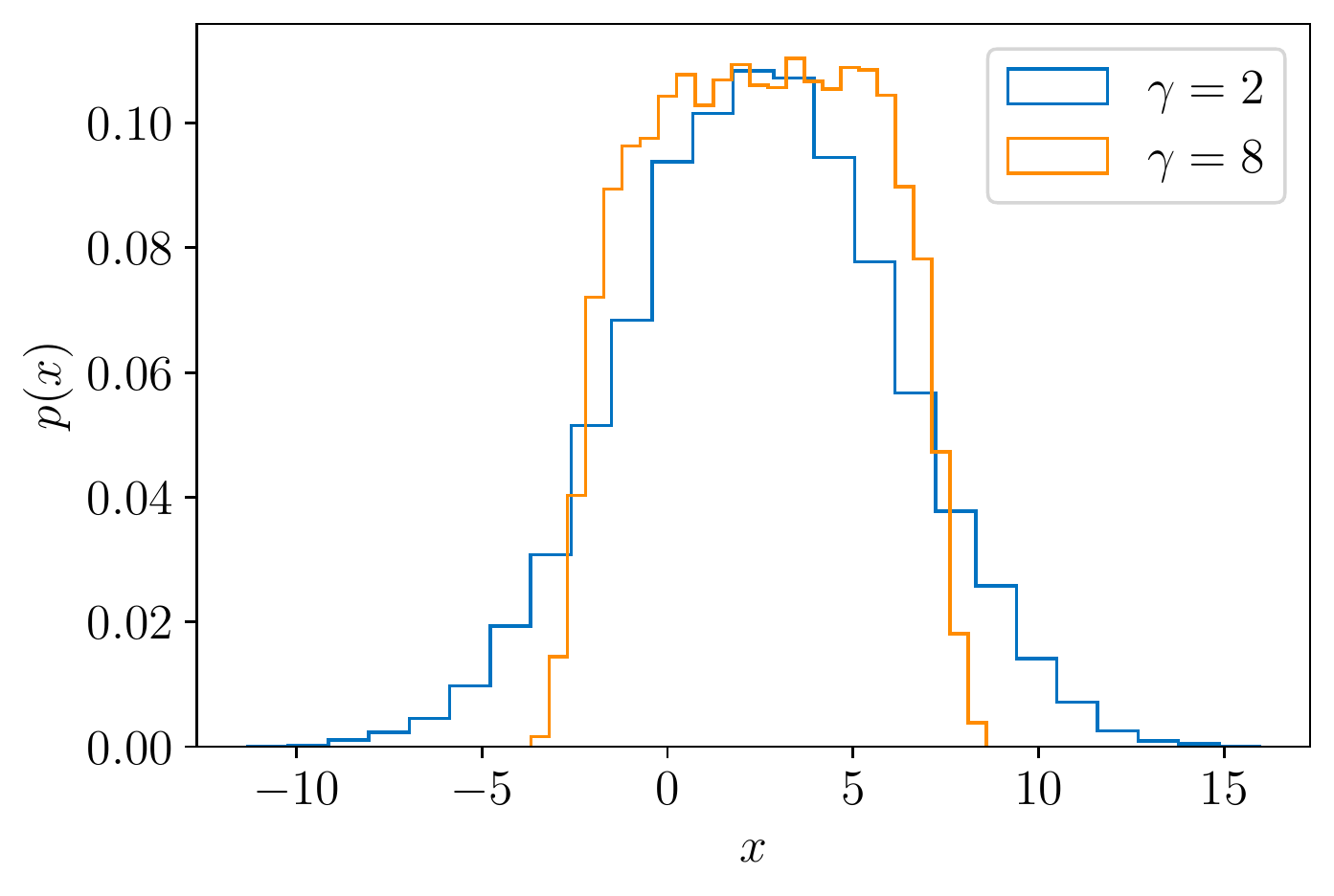}
    \caption{
    Realizations of samples from the standard Gaussian distribution (blue) and generalized Gaussian distribution with $\gamma=8$ (orange).
    The $\gamma=8$ case is closer to a tophat function that the standard Gaussian.
    }
    \label{fig:toy-model-data}
\end{figure}

As an example of these two distributions, we show the data used in the two examples considered in this section in Figure~\ref{fig:toy-model-data}.
In blue we show samples from the standard normal distribution, while the orange shows samples from the generalized distribution with $\gamma = 8$.
Thus, if the data follow a distribution with $\gamma \neq 2$, the value of $\alpha$ will be incorrectly estimated.
In the remainder of this section, we verify that our hybrid sampling method can recover $\mu$, $\alpha$, and $\gamma$ when we correctly assume that the data follows a standard Gaussian distribution during the first step of hybrid sampling and when the underlying data do not follow a Gaussian distribution.
For our analyses, we use prior distributions as described in Table~\ref{tbl:gaussian_priors}.

\begin{center}
\begin{table}
\begin{tabular}{|c c|}
 \hline
 Parameter & Distribution \\ 
 \hline
 $\mu$ & ${\cal U}(0, 5)$ \\
 $\alpha$ & ${\cal U}(0, 10 \sqrt{2})$ \\ 
 $\gamma$ & ${\cal U}(0, 10)$ \\
 \hline
\end{tabular}
\caption{
    Prior distributions for the parameters of the generalized Gaussian model.
    We denote a uniform distribution over $[a, b]$ as ${\cal U}(a, b)$.
    We note that the initial phase of the hybrid sampling fixes $\gamma=2$.
}\label{tbl:gaussian_priors}
\end{table}
\end{center}

\subsection{Well-Specified Initial Model} \label{sec:toy_well-specified}

\begin{figure}
    \centering
    \includegraphics[width=\linewidth]{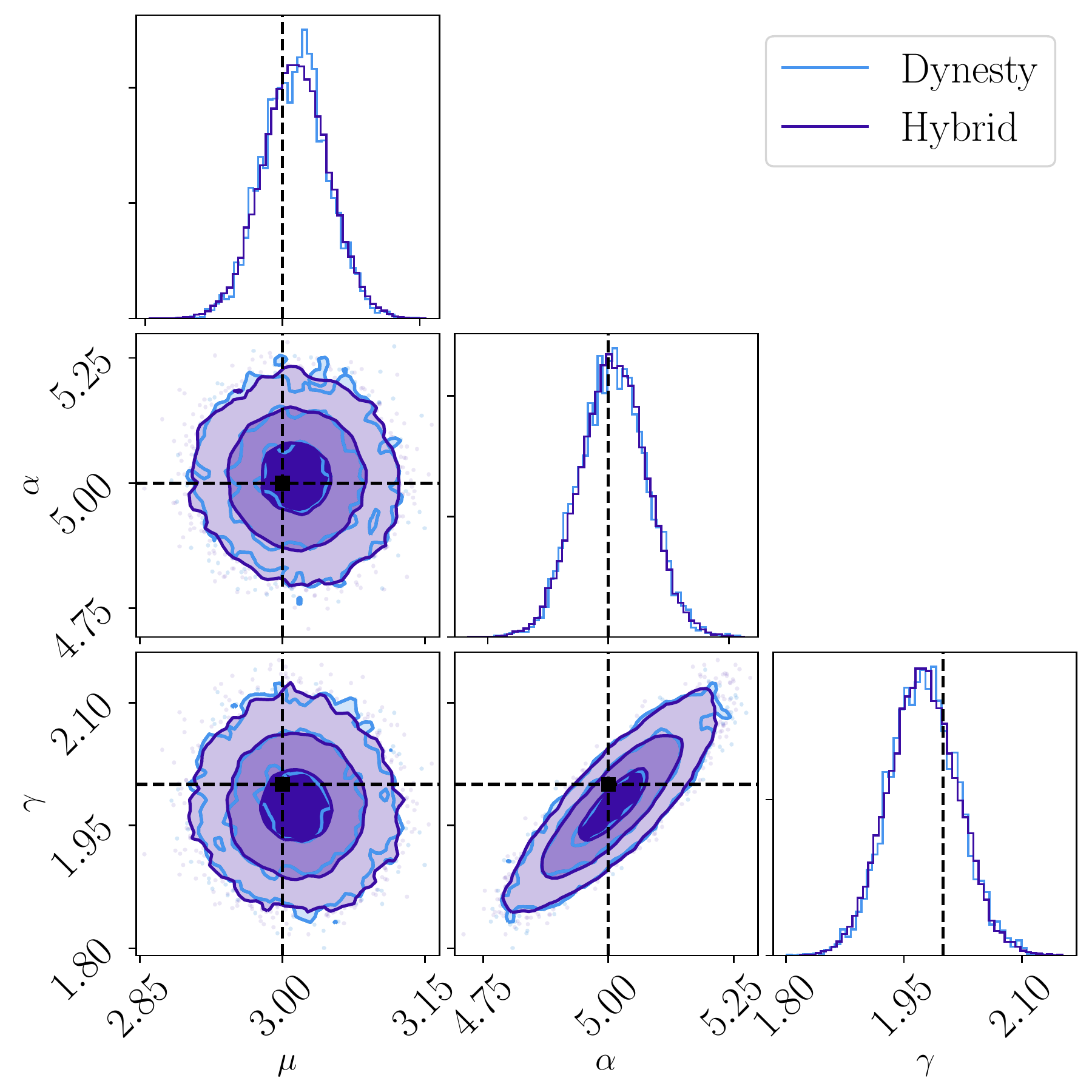}
    \caption{
        Posteriors for $\mu, \alpha, \gamma$ generated by our hybrid sampling method with the model well-specified during the first step, shown in purple, and \code{dynesty} sampling under a generalized Gaussian distribution for verification, shown in blue.
        True parameter values are shown in black.
        Contour level curves in this figure and all following two-dimensional distribution plots denote, from inside out, 39.3\% (1-sigma level), 86.4\% (2-sigma), and 98.8\% (3-sigma) of the distribution volume.
    }
    \label{fig:well-specified_comparison}
\end{figure}

First, we verify that hybrid sampling can recover model parameters when the data follows a normal distribution with $\mu=3$ and $\alpha=5$.
Our data $d$ are $N=10000$ random samples from this distribution.
In the first step of hybrid sampling, we use $\code{dynesty}$ to sample in $\{ \mu, \alpha \}$, assuming that the data follows a standard Gaussian distribution, generating a posterior distribution we denote $p_1(\mu, \sigma | d)$ using 500 live points.
Before the second step of hybrid sampling, we prepare initial points $\{ \mu_0, \alpha_0 \}$ for an ensemble of 200 walkers at seven temperatures as described in Section~\ref{sec:hybrid-sampling}.
We generate initial values of the shape parameter $\gamma_0$ by sampling from a standard Gaussian distribution with a standard deviation of $0.01$ and centered on the value of $\gamma$ assumed in the first hybrid step, $\gamma = 2$.
We then evolve the ensemble using \code{ptemcee} for 128 iterations, discarding the first 100 iterations as burn-in.

For comparison, we also analyze the data under the generalized model with \code{dynesty} directly.
In Figure \ref{fig:well-specified_comparison}, we compare the posterior generated by \code{dynesty} alone (blue) to the one generated by hybrid sampling (purple).
The dashed black lines show the true values of the three parameters.
We see that both methods recover equivalent posterior distributions indicating that the hybrid sampling method is well converged.

\subsection{Misspecified Initial Model} \label{sec:toy_misspecified}

\begin{figure}
    \centering
    \includegraphics[width=\linewidth]{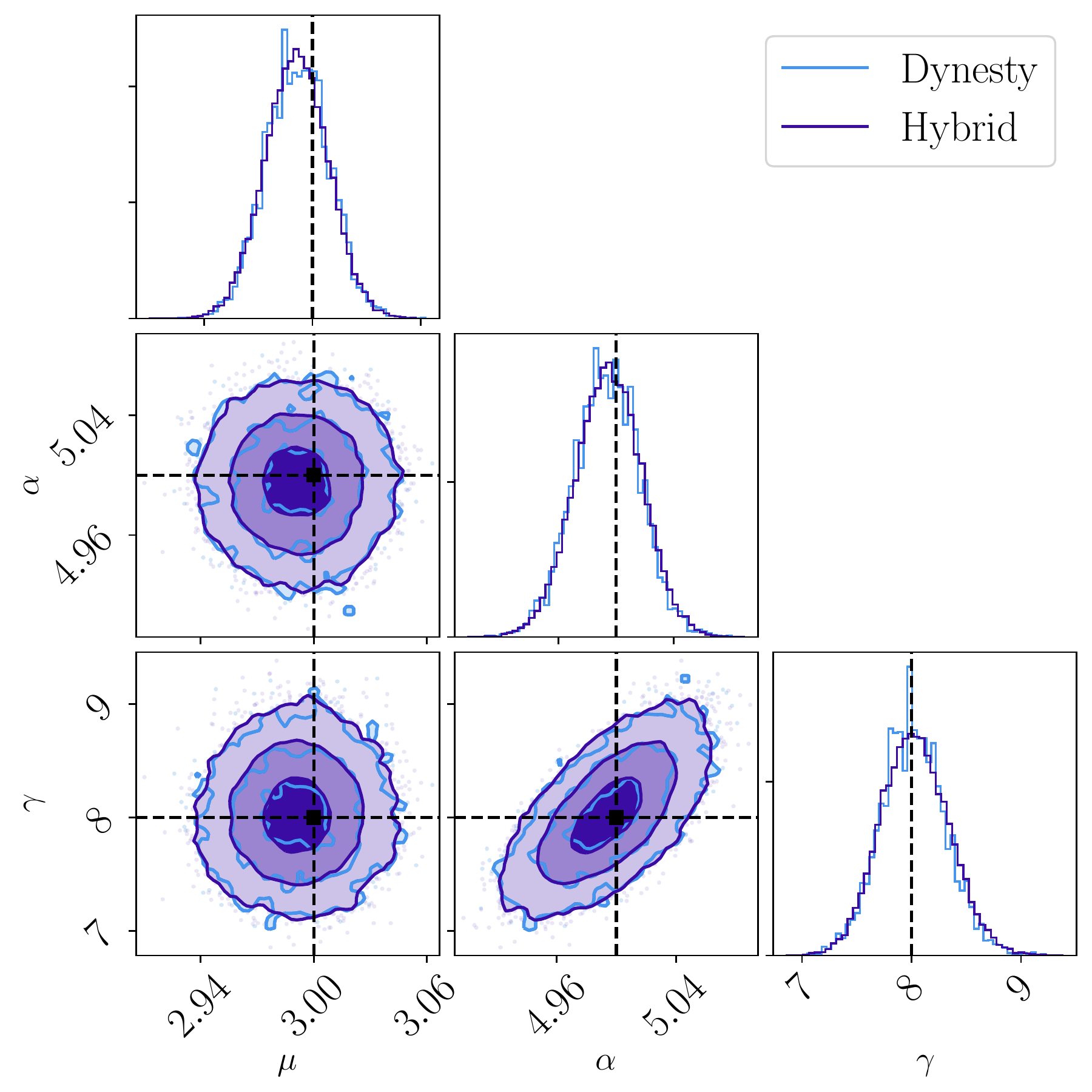}
    \caption{
        Posteriors for $\mu, \alpha, \gamma$ generated by our hybrid sampling method with the model misspecified during the first step, shown in purple, and \code{dynesty} sampling under a generalized Gaussian distribution for verification, shown in blue.
        True parameter values are shown in black.
    }
    \label{fig:misspecified_comparison}
\end{figure}

Next, we verify that hybrid sampling can recover model parameters when the model likelihood used in the first step has been inappropriately specified for the data.
We repeat the first step of hybrid sampling as in the previous section.
However, the input data we generate is not Gaussian.
Instead, we generate $N = 10000$ random samples from a generalized Gaussian distribution with $\mu = 3$, $\alpha = 5$, and $\gamma = 8$, shaped approximately like a tophat function.
We perform the same analyses as in Section~\ref{sec:toy_well-specified}, including the discard of the first 100 iterations as burn-in.
In Figure~\ref{fig:misspecified_comparison}, we show the posterior distributions estimated using our two methods.
Once again, we see that both methods recover equivalent results.

\begin{figure}
    \centering
    \includegraphics[width=\linewidth]{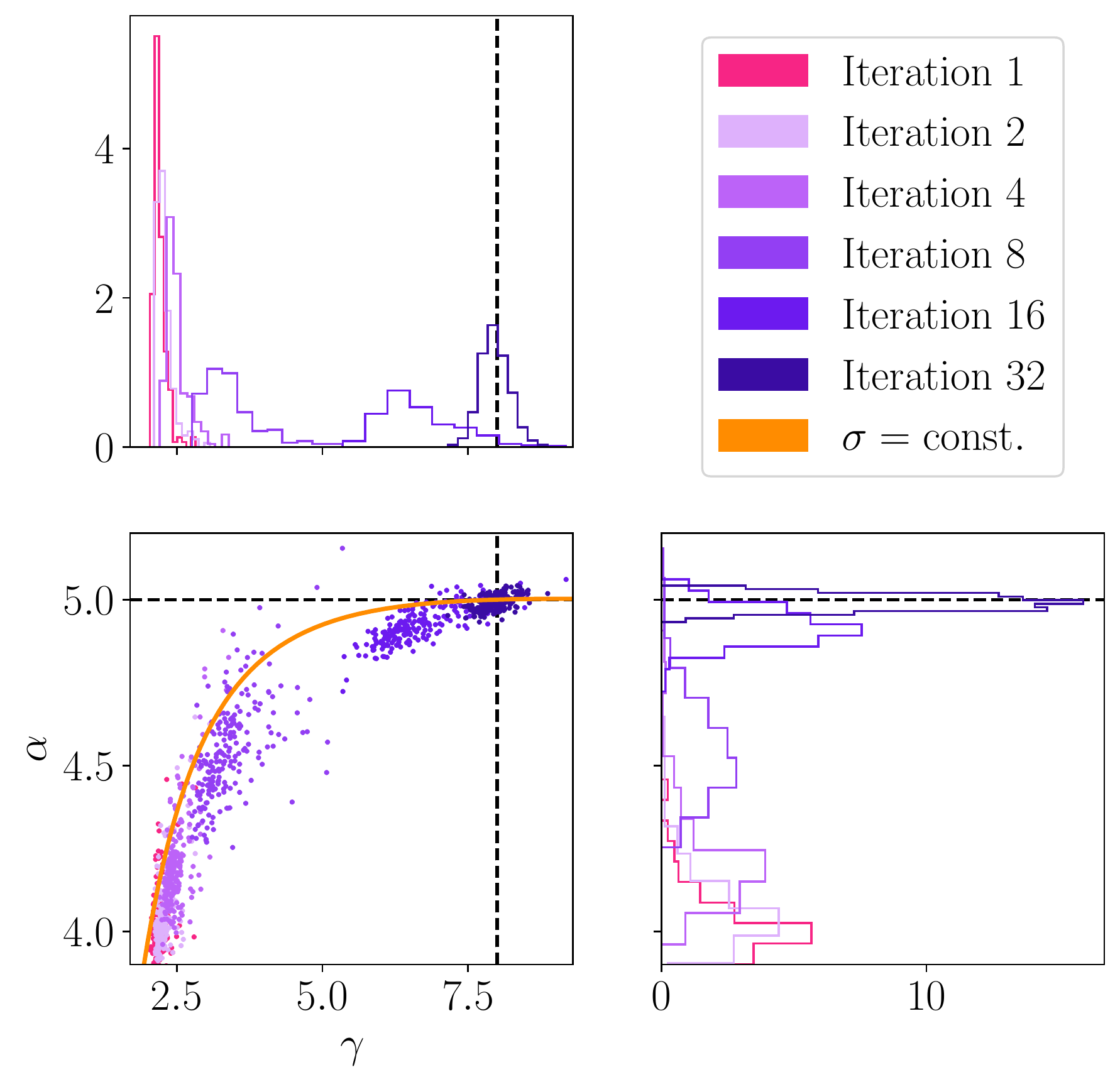}
    \caption{
        Visualization of the evolution of the $\beta_{T} = 1$ ensemble for our generalized Gaussian model during the MCMC sampling stage.
        The different colored scatter plots and histograms correspond to different iterations of the ensemble.
        The orange curve shows a line of constant $\sigma$ (Eq.~\ref{eq:sigma}) which describes the degeneracy between the $\alpha$ and $\gamma$.
        We note that the ensemble approximately evolves along a curve of constant $\sigma$.
        We show the full one-dimensional evolution in Figure~\ref{fig:misspecified_trace}.
    }
    \label{fig:alpha-beta_evolution}
\end{figure}

\begin{figure}
    \centering
    \includegraphics[width=\linewidth]{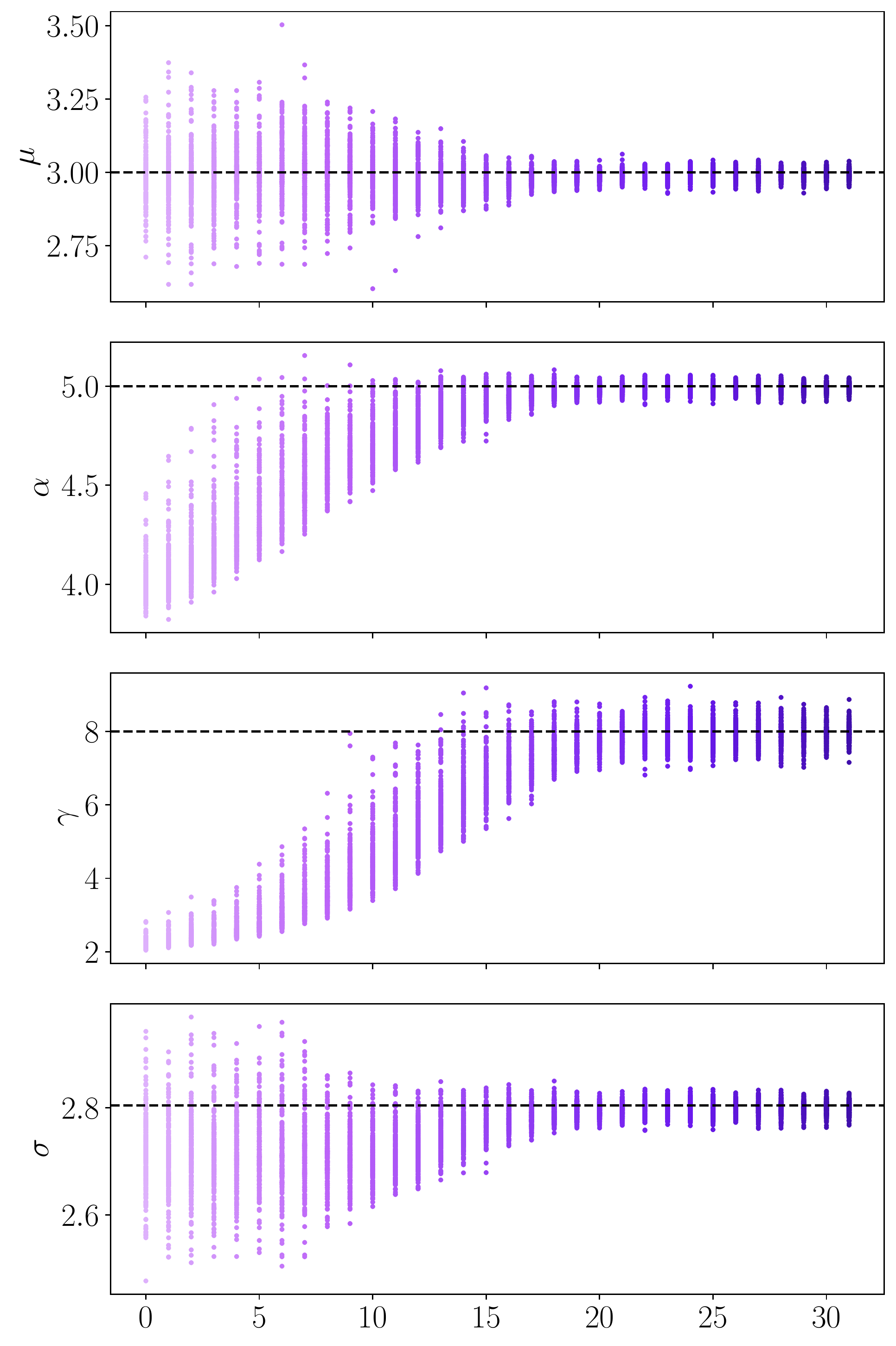}
    \caption{
        Trace plots showing the evolution of samples taken in $\mu$, $\alpha$, $\gamma$, and $\sigma$ during the second step of hybrid sampling for the first 32 steps of sampling, with black lines denoting the true values of these parameters.
        The samples plotted at each iteration of sampling are collated from each of the 200 walkers in the ensemble at temperature $\beta_T = 1$.
        The color scheme matches the state of the ensemble shown in Figure~\ref{fig:alpha-beta_evolution}.
        Throughout sampling, we observe that $\gamma$ and $\alpha$ achieve the correct values while not biasing our correct estimate of $\mu$ from the first sampling step. Additionally, we see that the convergence of $\sigma$ follows the convergence of the ensemble state in Figure~\ref{fig:alpha-beta_evolution} around a line of constant $\sigma$.
    }
    \label{fig:misspecified_trace}
\end{figure}

We are additionally interested in understanding the performance of the hybrid sampling stage.
In Figure~\ref{fig:alpha-beta_evolution} we provide two-dimensional snapshots from the evolution.
The top and left-hand panels show the one-dimensional marginal distributions for $\gamma$ and $\alpha$ at each iteration.
The colors are consistent between the panels and darker shades correspond to later iterations of the evolution.
We note that $\alpha$ and $\gamma$ are strongly correlated, specifically the variance of the distribution
\begin{equation}\label{eq:sigma}
    \sigma^2 = \alpha^2 \frac{\Gamma\left(\frac{3}{\gamma}\right)}{\Gamma\left(\frac{1}{\gamma}\right)}
\end{equation}
is approximately conserved.
The orange curve is constant $\sigma$ intersecting the true values of $\alpha$ and $\gamma$.
We note that the parameters $\alpha$ and $\gamma$ are correlated and so the evolution follows the direction of the correlation.
We observe that as the ensemble evolves, it follows this curve of constant $\sigma$.
This is suggestive that ensemble sampling can readily explore problems when the extended parameter space is strongly correlated with the initial parameter space.

In Figure~\ref{fig:misspecified_trace}, we show trace plots for $\mu$, $\alpha$, $\gamma$, and $\sigma$ from the second step of hybrid sampling.
At each iteration, we show the current state of the $\beta_{T}$ ensemble.
The color at each iteration matches the colors in Figure~\ref{fig:alpha-beta_evolution}.
The dashed lines indicate the true value of each parameter.
We see that in this case, the ensemble evolves from the initial state to the stationary distribution containing the true value within $<100$ iterations.

\section{Gravitational Wave Source Model} \label{sec:astro_methods}

\subsection{Modeling Gravitational Waveforms from Black Hole Mergers}

To infer the properties of the source of a gravitational-wave signal, we require a model for the gravitational waveform.
A quasi-circular binary black hole (BBH) merger can be described by 15 source parameters, divided between eight ``intrinsic'' parameters (the masses and spins of the component black holes) and seven ``extrinsic'' parameters (the location and orientation of the source with respect to an observer).
As these parameters describe the signal predicted by general relativity, we refer to these as ``GR parameters", later denoted $\theta_{\mathrm{GR}}$.
When modeling the signal from a binary black hole merger, the coalescence is typically broken down into three temporally-distinct regimes: the \textit{inspiral}, an \textit{intermediate} phase and the \textit{merger-ringdown} \cite{GWTC2_gr-tests_LIGO_2020,IMRPhenomPv2-II_khan2016,Agathos2014}.

The inspiral begins when the black holes have formed a binary system, however, we typically only model the waveform after the emission has surpassed the lowest sensitivity frequency of our instruments (typically $20$Hz for current detectors).
This regime is typically characterized using the post-Newtonian expansion \citep{blanchet_pn-review_2014} with higher-order corrections tuned to numerical relativity simulations.
During the intermediate regime, the orbital frequency of the binary increases to a point where the post-Newtonian expansion breaks down and the binary ``plunges'' and the horizons merge.
This regime can only be accurately described through numerical methods and is usually modeled using a fit to numerical relativity simulations (for example, public catalogs like \citep{sxs-catalog-2019}).
Finally, after the merger, the remnant black hole undergoes a ``ringdown'' phase, in which gravitational-wave emission is modeled via quasi-normal modes~\citep{qnm_gr-tests_review_berti_2009, ringdown-analysis-isi-2021}.
This regime is well-described by analytical models and provides strong tests of the ``no-hair'' theorem~\cite{gw150914-ringdown-tgr_isi_2019} and the black hole area law~\cite{Isi2021}.

In this work, we use \code{IMRPhenomPv2}, a computationally efficient, phenomenological model of the gravitational waveform~\citep{IMRPhenomPv2-I_husa2016, IMRPhenomPv2-II_khan2016, HannamIMRPhenomP, Bohe2016}.
For a given set of BBH source parameters, \code{IMRPhenomPv2} returns a frequency-domain representation of the gravitational wave signal, taking the form
\begin{equation}
    \tilde{h}(f) = A(f) e^{-i \Psi(f)}
\end{equation}
where $\tilde{h}(f)$ is the gravitational-wave strain as a function of frequency, $A(f)$ is the amplitude, and $\Psi$ is the phase of the signal.
Both $A$ and $\Psi$ depend on the intrinsic and extrinsic parameters of the BBH, although in general, the intrinsic parameters have a larger impact on the phase, while the extrinsic parameters primarily determine the amplitude.
In this work, we focus on modifications to the phase $\Psi$ as the first test of our hybrid sampling method for gravitational-wave signals, as current detectors are more sensitive to the phase of the signal~\citep{GWTC3_gr-tests_LIGO_2021}.

During the inspiral regime, $\Psi$ is approximated as a modified version of the post-Newtonian expansion: 
\begin{align}
    \Psi_{\mathrm{ins}}(f) &= 2\pi f t_{c} - \phi_{c} -\frac{\pi}{4} + \frac{3}{128 \eta v^5} \sum_{i=0}^{7} \left(\varphi_i + \varphi_{iL}\ln v \right) v^i
    \\
    &+ \frac{1}{\eta} \left( \sigma_{0} + \sigma_{1} f + \frac{3}{4}\sigma_{2} f^{4/3} + \frac{3}{5}\sigma_{3} f^{5/3} + \frac{1}{2}\sigma_{4} f^{2} \right). \nonumber
\end{align}
Here $\eta = m_1 m_2 / (m_1 + m_2)^2$ is the symmetric mass ratio of the binary, $v = \left( \pi M f G c^{-3} \right)^{1/3}$ is the dimensionless orbital frequency of the system, the phase coefficients $\varphi_i$ are determined by the post-Newtonian expansion and the $\sigma_{j}$ are tuned to numerical relativity waveforms.
Terms $\varphi_{iL}$ are those post-Newtonian coefficients leading $\ln{v}$ at order $i$.
Both $\varphi_{i}$ and $\sigma_{j}$ depend on the intrinsic parameters of the source.
The parameters $t_{c}$ and $\phi_{c}$ are the coalescence time and the orbital phase at coalescence respectively.

In the intermediate phase, \code{IMRPhenomPv2} adopts the following form for $\Psi$,
\begin{equation}
    \Psi_{\mathrm{int}}(f) = \frac{1}{\eta} \left( \beta_0 + \beta_1 f + \beta_2 \log(f) - \frac{\beta_3}{3} f^{-3} \right)
\end{equation}
where $\beta_0$ and $\beta_1$ are chosen to require a smooth continuation of $\Psi$ in the coalescence time and phase from the inspiral to intermediate phases.
The parameters $\beta_2$ and $\beta_3$ depend on the intrinsic parameters of the source.

Finally, in the merger-ringdown phase, \code{IMRPhenomPv2} adopts another parameterized form for $\Psi$,
\begin{equation}
    \begin{aligned}
    \Psi_{\mathrm{MR}} = \frac{1}{\eta} \biggl[ \alpha_0 + \alpha_1 f - \alpha_2 f^{-1} + \frac{4}{3} \alpha_3 f^{3/4}
    \\
    + \alpha_4 \tan^{-1}\left( \frac{f - f_{\mathrm{RD}}}{f_{\mathrm{damp}}} \right) \biggr].
    \end{aligned}
\end{equation}
As with the intermediate phase, $\alpha_0$ and $\alpha_1$ are chosen so that $\Psi$ continues smoothly from the intermediate phase to the merger-ringdown phase, and $\alpha_{2-4}$ depend on the intrinsic parameters of the source.
The frequencies $f_{\mathrm{RD}}$ and $f_{\mathrm{damp}}$ describe the complex ringdown frequency and are computed from the mass and spin of the remnant black hole~\citep{Bohe2016}.
We note that the above discussion applies to the aligned-spin \code{IMRPhenomD} model; the \code{IMRPhenomPv2} phasing is obtained by ``twisting-up'' the \code{IMRPhenomD} phasing to account for precession of the orbital plane as described in~\cite{Bohe2016}.

\subsection{Generating Beyond-GR Waveforms} \label{sec:bgr-waveforms}

Following~\cite{Agathos2014, single-BBH_gr-tests_LIGO_2016, GWTC1_gr-tests_LIGO_2019, GWTC2_gr-tests_LIGO_2020}, we model deviations from general relativity as fractional corrections to the parameters described above, specifically, we define
\begin{equation}
    p^{\rm BGR} = (1 + \delta p) p^{\rm GR}
\end{equation}
for $p^{\rm GR} \in \{ \varphi_{1-4}, \varphi_{5L,6L}, \varphi_{6,7}, \alpha_{2-4}, \beta_{2,3} \}$.
Since under GR $\varphi_1=0$, we model $\delta \varphi_{1}$ as an absolute rather than fractional deviation.
We note that the parameters describing global phase and time shifts are not modified as any such modification is indistinguishable from changes in phase or time.
In principle, any combination of these parameters could have non-zero deviations, however, most previous analyses modify just a single parameter at a time.
We use the implementation of \code{IMRPhenomPv2} provided by \code{LALSuite} \cite{lalsuite} which allows us to apply these fractional deviations when generating our waveforms.

Although these fractional deviations can take any real value, we do not expect them to be large as at worst we expect the general relativistic description of gravitational wave emission to be ``mostly" correct.
Thus, we limit the space of allowed deviations by limiting the allowed deviation between a beyond-GR waveform we generate with $p_{i, \mathrm{BGR}}$ and the associated waveform generated with $p_i$.
We measure this deviation between waveforms via the overlap $\mathcal{O}$,
\begin{equation}
    \mathcal{O} = \max_{\phi_{c}} \frac{\left< \tilde{h}_{\mathrm{GR}}(f), \tilde{h}_{\mathrm{BGR}}(f) \right>}{ \sqrt{ \left< \tilde{h}_{\mathrm{GR}}(f), \tilde{h}_{\mathrm{GR}}(f) \right> \left< \tilde{h}_{\mathrm{BGR}}(f), \tilde{h}_{\mathrm{BGR}}(f) \right> } }.
\end{equation}
Here $\tilde{h}_{\mathrm{GR}}$ and $\tilde{h}_{\mathrm{BGR}}$ are the GR frequency-domain waveform and associated beyond-GR waveform with the same intrinsic parameters.
We maximize over the merger phase of the signal by taking the absolute value of the overlap (e.g.,~\cite{Allen2012}).
One can similarly maximize over the merger time.
However, as detailed in Appendix~\ref{app:time-maximization}, we find that an overlap cut maximized over the merger phase and time introduces sufficient flexibility that the GR parameters can deviate significantly from the corresponding value with no beyond-GR deviation.
Finally, $\langle \cdot, \cdot \rangle$ denotes a discrete inner product between the frequency-domain waveforms, weighted by the detector spectral power density, as,
\begin{equation}
    \left< \tilde{h}_{1}(f), \tilde{h}_{2}(f) \right> = \frac{4}{T} \sum_{i}^{N} \frac{ \tilde{h}_{1,i} \tilde{h}^{*}_{2,i} }{ S_i },
\end{equation}
between two generic frequency-domain waveforms $\tilde{h}_{1}$ and $\tilde{h}_{2}$, where $i$ enumerates $N$ discrete sampling frequencies spaced by $1 / T$.
In practice, we use the +-polarization of the waveform for computing the overlaps.
The quantity $S$ is the harmonic sum of the power spectral densities for each of the interferometers in the network.
In this work, for some cases of hybrid sampling, we enforce a cut on the priors of $\delta p_i$ by enforcing that all beyond-GR waveforms we generate must have an overlap $\mathcal{O} > 0.9$ with their associated GR waveform. This manifests in practice as a cut on the prior bounds of the initial points provided to \code{ptemcee}, as well as an added acceptance condition for MCMC proposals, where any proposal with $\mathcal{O} < 0.9$ is rejected.

\section{Hybrid Sampling in Gravitational Wave Signals} \label{sec:hybrid-sampling_gws}

We now apply our method to real and simulated gravitational-wave signals.
We follow the procedure described in Section~\ref{sec:stat_methods} to jointly infer $\theta_{\rm GR}$ and each of the $\delta p$ parameters.
For each analyzed signal, we first analyze the data using \code{dynesty} under the GR model.\footnote{We note that, in practice, this analysis is typically performed by the LIGO/Virgo/Kagra collaboration and so would not be required in production scenarios if the nested samples are released for future analyses.}
Unless otherwise specified, we then perform 28 subsequent analyses with \code{ptemcee}, two each allowing one of the $\delta p$ parameters to vary either applying the condition that ${\cal O} > 0.9$ or no overlap cut.
For all analyses, we numerically marginalize the likelihood over distance and the coalescence phase using standard methods~\cite{bayesian_inference_gws_thrane_2019}.
Full details of the sampler configurations can be found in Appendix~\ref{sec:sampler_settings}.

The prior distribution we use for $\theta_{\rm GR}$ is given in Table~\ref{tbl:gw150914_gr_priors}.
We note that throughout we work with detector-frame mass quantities which differ from the source mass by a distance-dependent factor due to cosmological redshifting.
For the \code{ptemcee} stage, we initialize the $\theta_{\rm GR}$ from the tempered posterior distribution obtained with \code{dynesty}.
The prior and initialization distributions for the $\delta p$ are shown in Table~\ref{tbl:gw150914_dpi_distributions}.

\begin{center}
\begin{table}
\begin{tabular}{|c c c|}
 \hline
 Parameter & Distribution & Unit \\ 
 \hline
 $m_{1}, m_{2}$ & ${\cal U}(1, 1000)$ & $ M_{\odot}$ \\
 $\mathcal{M}$ & $[21.418, 41.974]$ & $M_{\odot}$  \\ 
 $q$ & $[0.05, 1.0]$ & - \\
 $a_{1}$, $a_2$ & $\mathcal{U}(0, 0.99)$ & - \\
 $\theta_{1}, \theta_{2}, \theta_{\rm JN}, \kappa$ & Sin & rad\\
 $\phi_{12}, \phi_{jl}, \phi_{c}, \epsilon$ & $\mathcal{U}(0, 2\pi)$ & rad \\
 $\psi$ & $\mathcal{U}(0, \pi)$ & rad \\
 $t_c$ & ${\cal U}(t_{0} - 0.1, t_{0} + 0.1)$ & $s$ \\
 $d_L$ & ${\cal P}(2, 10, 10^4)$ & Mpc \\
 \hline
\end{tabular}
\caption{
    Prior distributions for $\theta_{GR}$ used in both steps of hybrid sampling to estimate the source properties of the gravitational-wave signals we consider.
    We denote a uniform distribution over $[a, b]$ as ${\cal U}(a, b)$, ${\cal P}(\alpha, a, b)$ is a power-law distribution with spectral index $\alpha$ over the same domain.
    The sine distribution for a quantity $x$ is equivalent to a uniform distribution of $\cos(x)$.
    The notation $[a,b]$ denotes a parameter that is constrained to lie within that interval with the functional form defined in terms of other parameters.
    The prior for the coalescence time is centered on either the trigger time from the matched filter search pipelines for GW150194 or the known injection time for simulated signals.
    Parameter definitions follow~\cite{RomeroShaw2020}.
}\label{tbl:gw150914_gr_priors}
\end{table}
\end{center}

\begin{center}
\begin{table}
\begin{tabular}{|p{3cm}  p{2cm} p{2cm}|}
 \hline
 Parameter & Prior & Initialization \\ 
 \hline
 $\delta \varphi_0$ & $\mathcal{U}(-1, 1)$ & $\mathcal{N}(0, 10^{-2})$ \\
 $\delta \varphi_1$ & $\mathcal{U}(-2, 2)$ & $\mathcal{N}(0, 10^{-1})$ \\
 $\delta \varphi_2, \delta \varphi_3, \delta \varphi_4, \delta \varphi_{5l}$ & $\mathcal{U}(-5, 5)$ & $\mathcal{N}(0, 1)$ \\
 $\delta \varphi_6$ & $\mathcal{U}(-10, 10)$ & $\mathcal{N}(0, 1)$ \\
 $\delta \varphi_{6l}, \delta \varphi_7$ & $\mathcal{U}(-30, 30)$ & $\mathcal{N}(0, 5)$ \\
 $\delta \alpha_2, \delta \alpha_3, \delta \alpha_4$ & $\mathcal{U}(-5,5)$ & $\mathcal{N}(0, 1)$ \\
 $\delta \beta_2, \delta \beta_3$ & $\mathcal{U}(-5, 5)$ & $\mathcal{N}(0, 1)$\\
 \hline
\end{tabular}
\caption{
    Prior (center) and initialization (right) distributions for the post-Newtonian deviation parameters $\delta p_i$ used in the \code{ptemcee} step of hybrid sampling for GW150914.
    The prior distributions were chosen to fully include the $\delta p_i$ posteriors for GW150914 in \cite{single-BBH_gr-tests_LIGO_2016}.
    The initialization distributions were chosen to be narrower than the expected posterior distributions.
    Here, ${\cal U}(a, b)$ denotes a uniform distribution in $[a, b]$ and ${\cal N}(\mu, \sigma)$ a normal distribution with mean $\mu$ and standard deviation $\sigma$.
}\label{tbl:gw150914_dpi_distributions}
\end{table}
\end{center}

\subsection{Analysis of a Real Signal - GW150914}
First, we apply our hybrid sampling method on GW150914, the first observed gravitational-wave signal \cite{gw150914_properties}.
This signal was produced by the coalescence of a binary black hole system with a detector-frame chirp mass of $\mathcal{M} \sim 30 M_{\odot}$ and a network signal-to-noise ratio of $\sim 25$.
These properties mean it is still one of the highest SNR signals to date and also lies at the mode of the observed binary black hole mass distribution~\cite{O3bPop} making it an excellent representative test case.
Following~\citep{gw150914_properties}, we analyze $8s$ of data ending $2s$ after the trigger time produced by matched-filter search pipelines for both of the Advanced LIGO interferometers.
We use the power spectral densities and calibration envelopes used in the LIGO/Virgo collaboration analyses available at~\cite{GWTC1Data}.
Marginalizing over uncertainty in the detector calibration adds 40 free parameters to the analysis, and we use the same prior distribution for those parameters as~\cite{GWTC1}.
We downsample the data to $2048$ Hz and analyze the data from $20-1024$ Hz.

\begin{figure*}
    \centering
    \includegraphics[width=\linewidth]{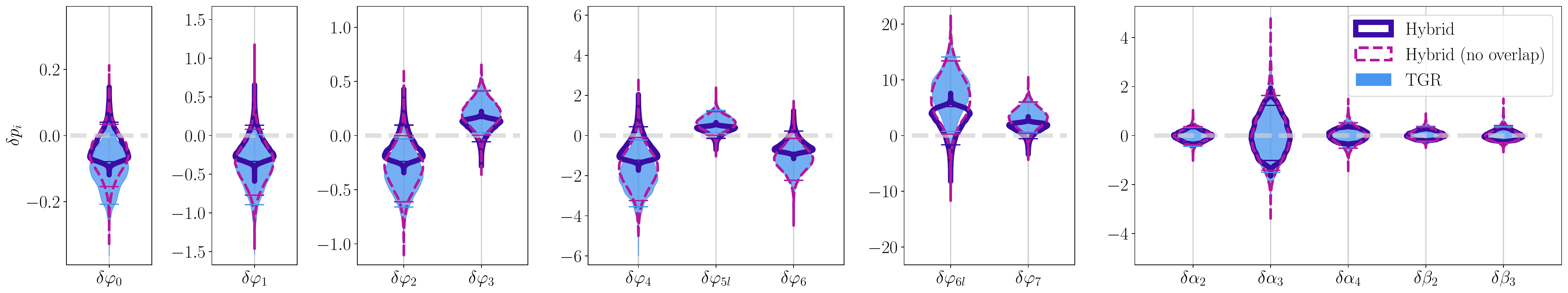}
    \caption{
        Violinplot, showing the posterior distributions on each post-Newtonian deviation parameter $\delta p_i$, comparing the results of \cite{GWTC1_gr-tests_LIGO_2019} with the results of our hybrid sampling method in the ``overlap cut" (purple, $\mathcal{O} > 0.9$) and ``no overlap cut" (magenta) cases.
        These results were obtained through 14 independent analyses, in each case, we vary only one deviation parameter at a time.
        Colored horizontal bars denote the 5th and 95th percentiles of the posteriors.
    }
    \label{fig:gw150914_violinplot}
\end{figure*}

In Figure~\ref{fig:gw150914_violinplot}, we show the posterior distributions for each of the $\delta p$ obtained with (purple) and without (magenta) a cut in the GR vs. non-GR overlap respectively.
We also overlay the results from the LIGO/Virgo collaboration analysis in blue obtained with \code{LALInference}~\cite{Veitch2015,GWTC1_gr-tests_LIGO_2019}.
The differences between the blue and magenta are likely due to sampler differences.

\begin{figure}
    \centering
    \includegraphics[width=\linewidth]{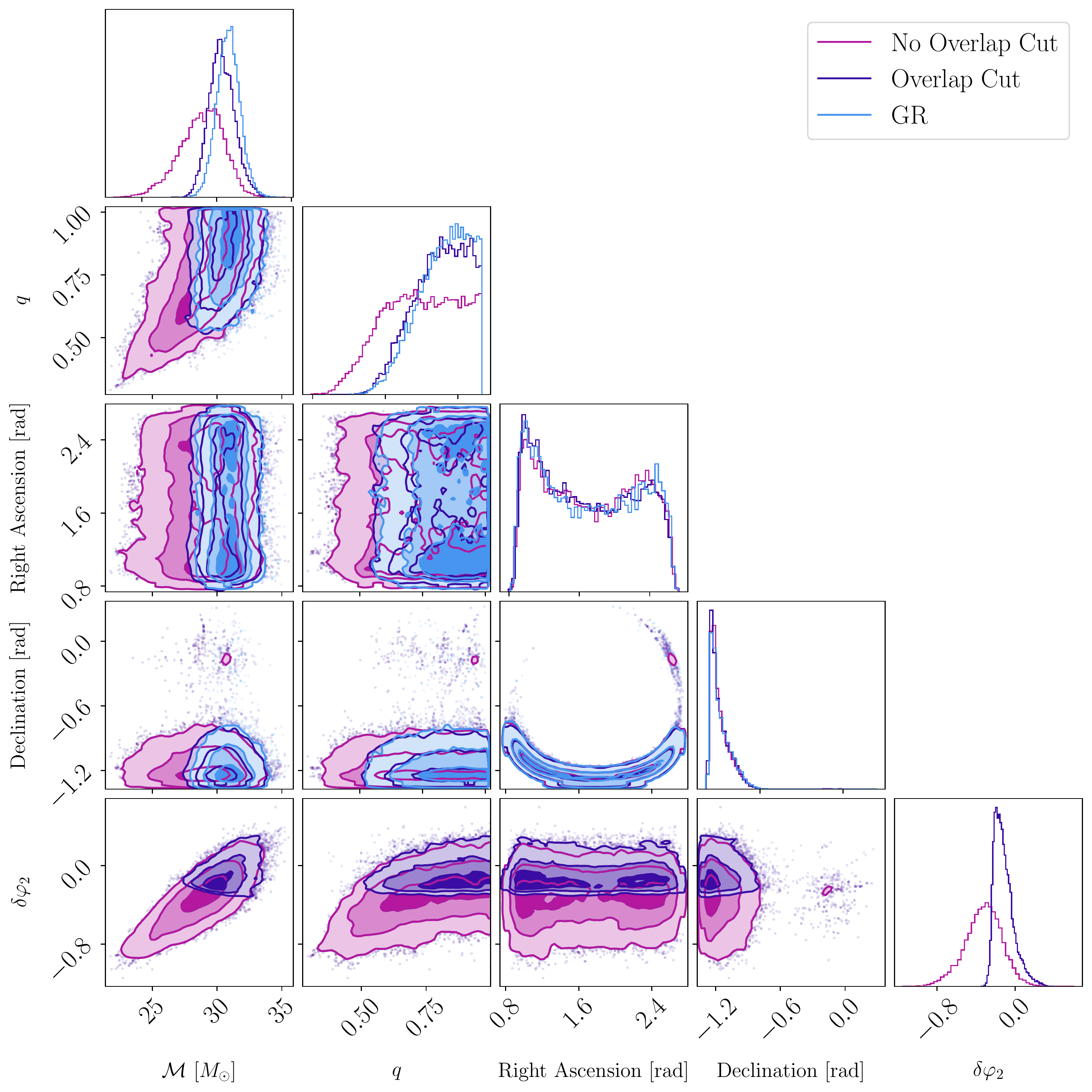}
    \caption{Joint posterior distributions on chirp mass $\mathcal{M}$, mass ratio $q$, right ascension, declination, and deviation $\delta \varphi_2$ during our estimation of $\delta \varphi_2$ in GW150914 with hybrid sampling. In magenta, we plot the posteriors with no overlap cut enforced, whereas in purple we enforce an overlap cut of $\mathcal{O} > 0.9$. In blue, we show posteriors for $\mathcal{M}$ and $q$ from the first step of hybrid sampling, with no deviations from general relativity. In both hybrid results, we see a correlation between $\delta \varphi_2$ and the intrinsic parameters, particularly $\mathcal{M}$, and no correlation between $\delta \varphi_2$ and the extrinsic sky parameters. We also observe the prior boundary imposed by the overlap cut that prevents our sampler from exploring values of the mass parameters more distant from the initial, general relativity-only result. }
    \label{fig:gw150914_dphi2_overlap-comparison}
\end{figure}

We note that for the inspiral deviation parameters the requirement that ${\cal O} > 0.9$ imposes a significant constraint compared to the constraining power of the data.
This is because the inspiral deviation parameters are strongly degenerate with the chirp mass, and also show correlations with the mass ratio as can be seen in Figure~\ref{fig:gw150914_dphi2_overlap-comparison}.
However, for the $\delta \alpha$ and $\delta \beta$ parameters, the posteriors are unaffected by the requirement that ${\cal O} > 0.9$.
This is because these parameters are not strongly correlated with the GR parameters and so an equivalent waveform cannot be obtained by changing, e.g., the black hole masses and $\delta \alpha_{2}$.

In Figure~\ref{fig:gw150914_dphi2_overlap-comparison}, we show joint posterior distributions on the beyond-GR deviation parameters $\delta \varphi_2$, intrinsic parameters chirp mass $\mathcal{M}$ and mass ratio $q$, and the extrinsic sky parameters right ascension and declination from our estimation of $\delta \varphi_2$ when enforcing an overlap cut of $\mathcal{O} > 0.9$ (purple) as well as enforcing no overlap cut (magenta).
We also compare these distributions to the posteriors in $\mathcal{M}$, $q$, right ascension, and declination generated during the first step of hybrid sampling, where we do not yet sample in deviations from general relativity. 
From the construction of the post-Newtonian inspiral phase coefficients, we expect deviations from $\varphi_2$ to be correlated with changes in the mass parameters, particularly $\mathcal{M}$, and we can observe this correlation in both results.
Since the extrinsic parameters do not affect the phase evolution of the signal, we do not expect a correlation between $\delta \varphi_{2}$ and the extrinsic parameters.
As expected, we do not see a correlation between $\delta \varphi_2$ and the extrinsic sky parameters.
We also observe the effect of the overlap cut, which prevents our ensemble from exploring far away from the GR solution for the mass parameters.

\begin{figure}
    \centering
    \includegraphics[width=\linewidth]{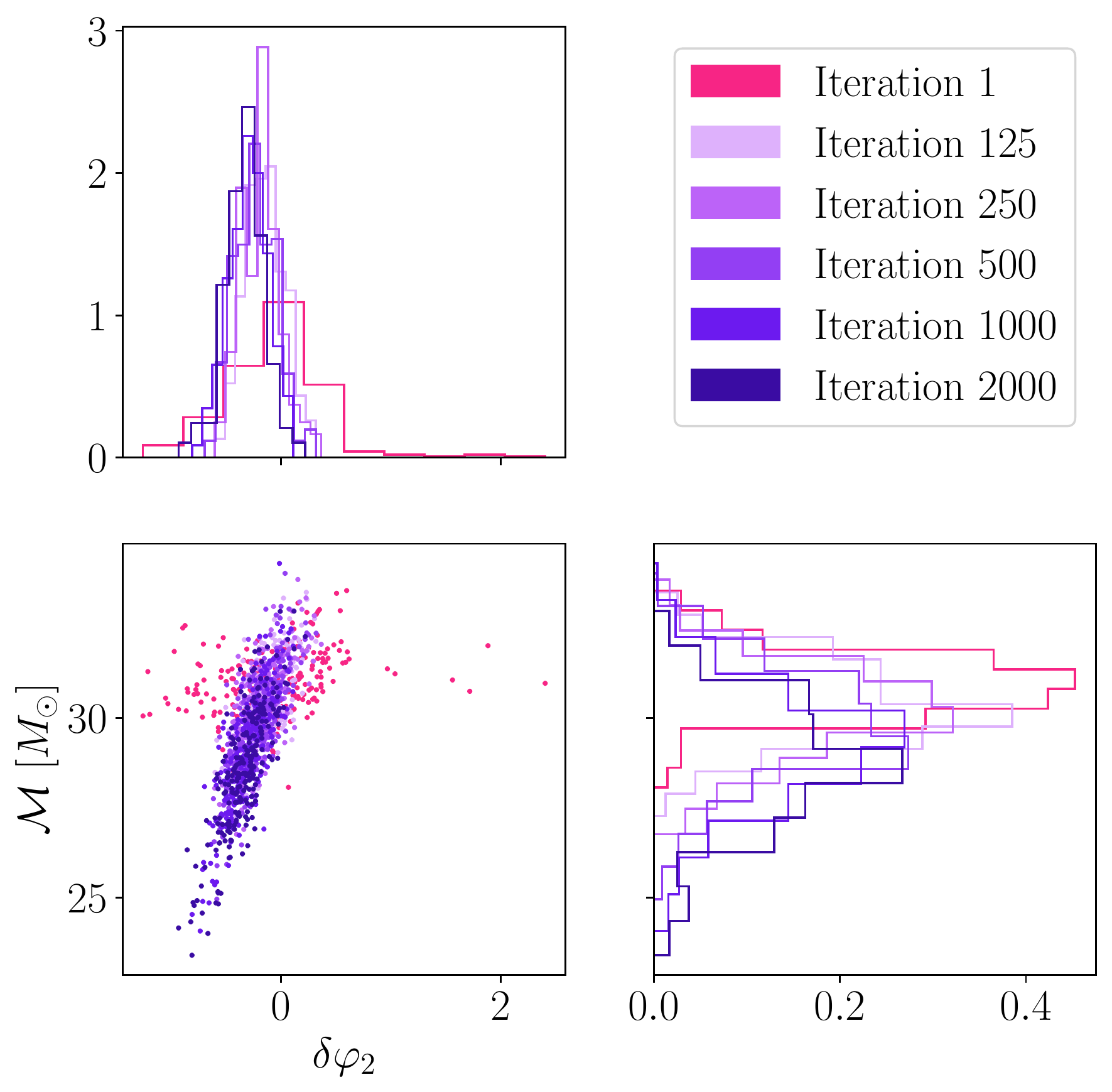}
    \caption{
        Snapshots of the $\beta_{T} = 1$ ensemble of our parallel-tempered ensemble MCMC analysis at various iterations for our analysis of GW150914 allowing the inspiral deviation parameter $\delta\varphi_{2}$ to vary. 
        We display chirp mass $\mathcal{M}$ against inspiral phase deviation coefficient $\delta\varphi_2$ with marginal distributions for $\mathcal{M}$ in the right column and those for $\delta\varphi_2$ in the top row. 
        In pink, we show the state of the ensemble after the first MCMC step, near its initialization from the posterior generated with \code{dynesty}. 
        As the ensemble evolves, represented with darkening shades of purple, the posterior expands to fill the extended posterior space.
        We note that the analysis correctly captures the expected correlation between the two parameters.
        In this analysis, we do not apply any condition on the overlap between the beyond-GR waveform and the corresponding GR waveform.
    }
    \label{fig:gw150914_evolution_no-overlap}
\end{figure}

In Figure~\ref{fig:gw150914_evolution_no-overlap}, we examine the evolution of the ensemble sampler for our analysis allowing $\delta \varphi_{2}$ to vary with no minimum allowed overlap.
We show the distribution of $\mathcal{M}$ and
$\delta \varphi_2$ at various iterations of the \code{ptemcee} analysis.
As in Section~\ref{sec:toy_misspecified}, the hybrid analysis method is well able to capture the correlation between chirp mass and the new parameter added in the second stage of our hybrid analysis.
We find that, by iteration 1024, the ensemble has converged to the correct solution.
In Appendix~\ref{app:150914-traces}, we provide additional plots showing the evolution of the $\beta_{T}=1$ ensemble for each of the analyses of GW150914.
In general, we find $\sim 1000$ iterations are sufficient to ensure convergence of the algorithm.

\begin{center}
\begin{table}
\begin{tabular}{|c | r|}
 \hline
 Parameter & $n_{\rm likelihood}$ \\ 
 \hline
 GR & 23,200,000 \\
 $\delta \varphi_0$ & 2,955,000 \\
 $\delta \varphi_1$ & 2,952,500 \\
 $\delta \varphi_2$ & 2,952,500 \\
 $\delta \varphi_3$ & 2,952,500 \\
 $\delta \varphi_4$ & 2,911,250 \\
 $\delta \varphi_{5l}$ & 2,951,250 \\
 $\delta \varphi_6$ & 2,953,750 \\
 $\delta \varphi_{6l}$ & 2,951,250 \\
 $\delta \varphi_7$ & 2,490,000 \\
 $\delta \alpha_2$ & 2,951,250 \\
 $\delta \alpha_3$ & 2,952,500 \\
 $\delta \alpha_4$ & 2,951,250 \\
 $\delta \beta_2$ & 2,951,250 \\
 $\delta \beta_3$ & 2,952,500 \\
 \hline
\end{tabular}
\caption{
    The number of likelihood evaluations required to estimate $\delta p_i$ in GW150914 using hybrid sampling.
    For reference, we include the number of likelihood evaluations required for the initial GR analysis.
    With the (optimistic) assumption that performing nested sampling to infer the $\delta p_{i}$ requires the same number of likelihood evaluations as with the GR model, our method is $\sim 8\times$ more efficient.
} \label{tbl:gw150914_cost}
\end{table}
\end{center}

We now assess whether our hybrid sampling method is more computationally efficient than the previously employed direct sampling method.
To do this, we compare the number of likelihood evaluations needed to produce well-converged results.
The computational cost for hybrid sampling scales linearly with the number of extensions to the base model.
A fixed number of likelihood evaluations are necessary for the first step of sampling with \code{dynesty}, followed by additional evaluations for each second step analysis performed with \code{ptemcee}.
Using \code{dynesty} alone in a ``standard" methodology also scales linearly without an initial fixed cost but, in general, each analysis with \code{dynesty} is more expensive than the same second-step hybrid analysis.
Thus, if we only seek to estimate a small number of $\delta p_i$, using \code{dynesty} alone may be more efficient, but we expect hybrid sampling to be more efficient after some break-even number of deviation parameter estimations.

We summarize the computational cost of each of the analyses we performed for our analysis of GW150914 in Table~\ref{tbl:gw150914_cost}.
For the initial GR-only inference we required 23.6 million likelihood evaluations and each \code{ptemcee} analysis required $< 3$ million likelihood evaluations.
We don't have access to the computational cost for the LIGO/Virgo analysis, however, we can conservatively estimate that direct \code{dynesty} sampling for each non-GR parameter will be at least as expensive as the GR-only analysis.
Additionally, in Appendix~\ref{app:150914-direct-ptemcee}, we directly sample in the GR parameters plus one non-GR parameter using \code{ptemcee} initialized at the maximum likelihood point, as well as with random samples from the prior, and find that these analyses had not converged after $> 6$ million likelihood evaluations.
We can therefore estimate that our hybrid sampling scheme is between $\sim 2$ and $\sim 10\times$ more efficient than the direct MCMC sampling method for this event.

\subsection{Simulated Non-GR Signals}

Analyses of real gravitational-wave transients have not revealed significant deviations from relativity, however, it is important to test whether our method will be sensitive to such effects if they are present.
To accomplish this, we analyze a simulated signal with a non-zero value of $\delta \varphi_{2}$ with our hybrid method; the specific injection parameters are described in Table~\ref{tbl:injected_parameters}.
We add this signal to the Advanced LIGO Livingston and Hanford interferometers assuming their design sensitivities~\cite{observing_scenarios} resulting in an injection with a network signal-to-noise ratio ${\rm SNR} \approx 370$.

\begin{center}
\begin{table}
\begin{tabular}{|c | c c|}
 \hline
 Parameter & Value & Unit \\
 \hline
 $\mathcal{M}$ & $30$ & $M_{\odot}$ \\
 $q$ & 0.8 & - \\
 $a_1, a_2$ & 0 & - \\
 $\theta_1$, $\theta_2, \phi_{12}, \phi_{jl}, \theta_{\rm JN}, \phi_c, \psi$ & 0 & rad \\
 right ascension & 1.35 & rad \\
 declination & -1.21 & rad \\
 $\delta \varphi_2$ & 0.2 & - \\
 other $\delta p$ & 0 & - \\
 $t_c$ & 0 & s \\
 $d_L$ & 100 & Mpc \\
 \hline
\end{tabular}
\caption{
    Parameters of the simulated signal injected into the Advanced LIGO gravitational wave detector network.
} \label{tbl:injected_parameters}
\end{table}
\end{center}

We follow the same hybrid sampling procedure as in our analysis of GW150914, and with the sampler settings found in Appendix \ref{sec:sampler_settings}, Table \ref{tbl:injected_sampler_kwargs}.
We also perform the beyond-GR analyses with \code{dynesty} without imposing an overlap cut, to compare the results between the two methods.

\begin{figure}
    \centering
    \includegraphics[width=\linewidth]{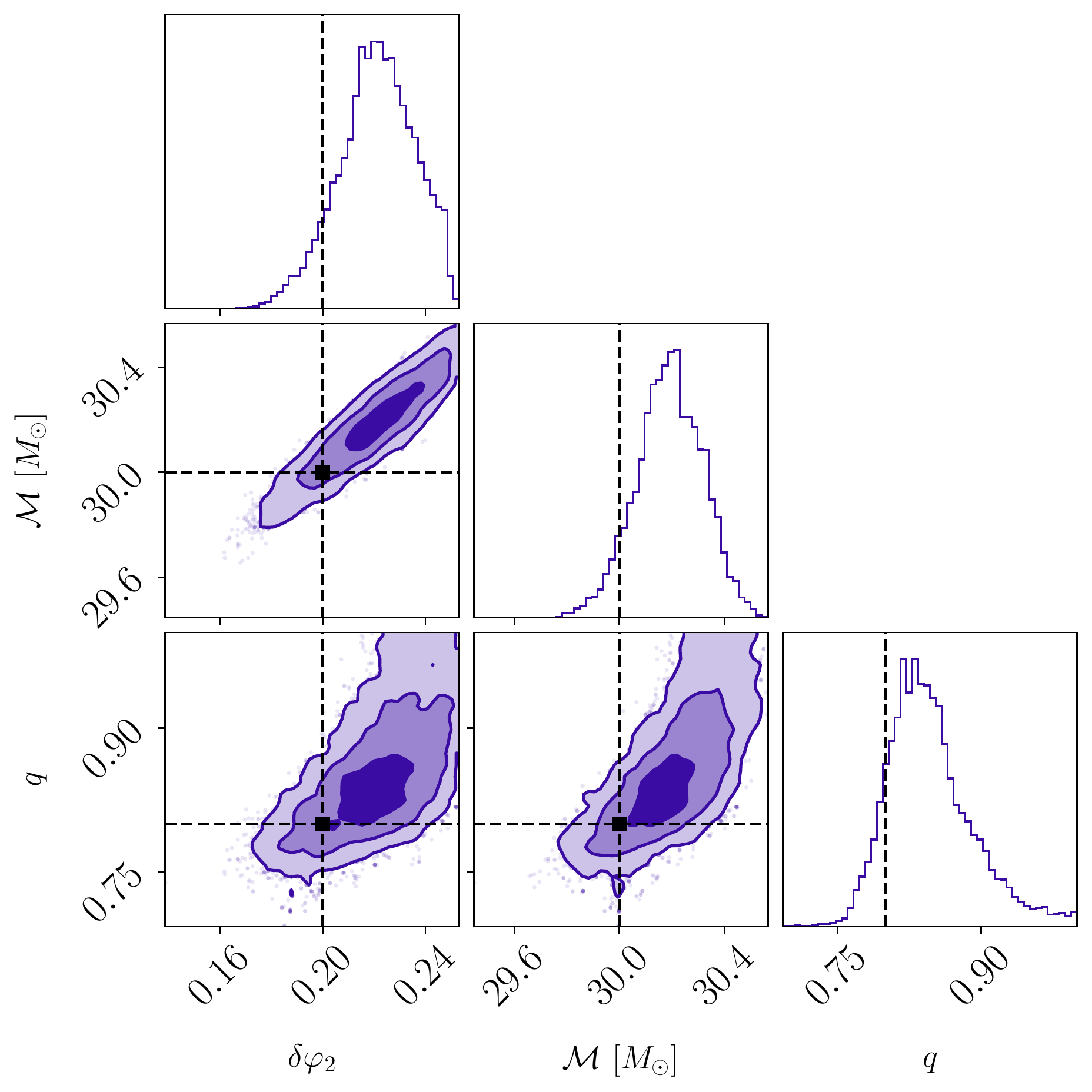}
    \caption{
        Corner plot, displaying marginal and joint posterior distributions on the inspiral regime deviation parameter $\delta \varphi_2$, chirp mass $\mathcal{M}$, and mass ratio $q$ from our injected signal generated by hybrid sampling with an overlap cut of $\mathcal{O} > 0.9$. As in Figure~\ref{fig:gw150914_dphi2_overlap-comparison}, we note that $\delta \varphi_2$ is correlated with the intrinsic mass parameters.
    }
    \label{fig:highsnr_corner}
\end{figure}

\begin{figure}
    \centering
    \includegraphics[width=\linewidth]{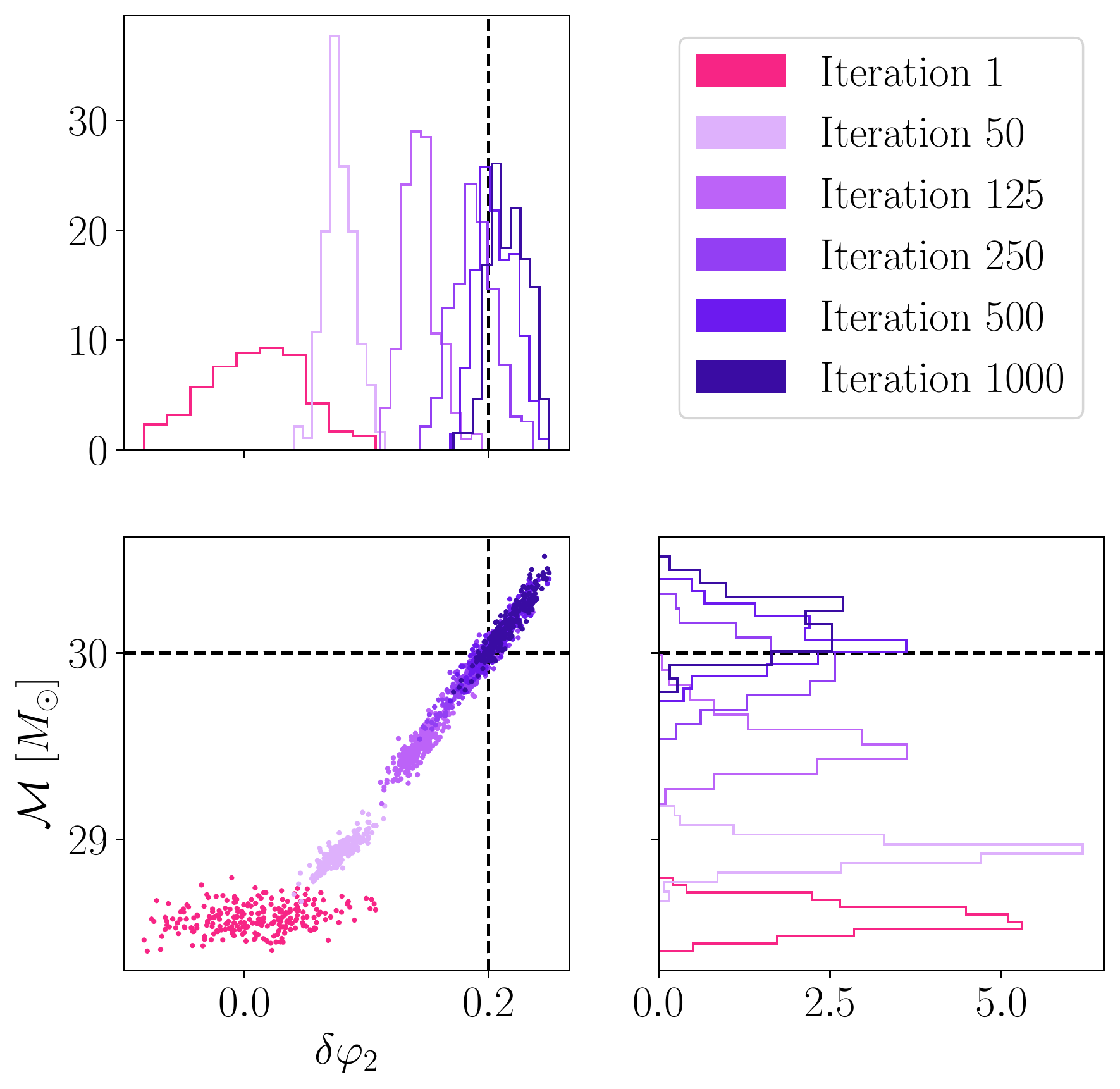}
    \caption{
        We plot the evolution of the second step of our hybrid analysis of the injected signal, with an overlap cut of $\mathcal{O} > 0.9$ imposed on waveform generation.
        We display chirp mass $\mathcal{M}$ against inspiral phase deviation coefficient $\delta \varphi_2$ with marginal distributions for $\mathcal{M}$ in the right column and those for $\delta \varphi_2$ in the top row. 
        In pink, we show the state of the ensemble after the first MCMC step, near its initialization from the posterior generated with \code{dynesty}. 
        As the ensemble evolves, shown with darkening shades of purple, it evolves a tight correlation between $\mathcal{M}$ and $\delta \varphi_2$ at a roughly constant overlap.
        The ensemble converges to correct values for $\delta \varphi_2$ and $\mathcal{M}$, inconsistent with the initial estimate of $\mathcal{M}$ from the first step of hybrid sampling.
    }
    \label{fig:highsnr_evolution}
\end{figure}

\begin{figure*}
    \centering
    \includegraphics[width=\linewidth]{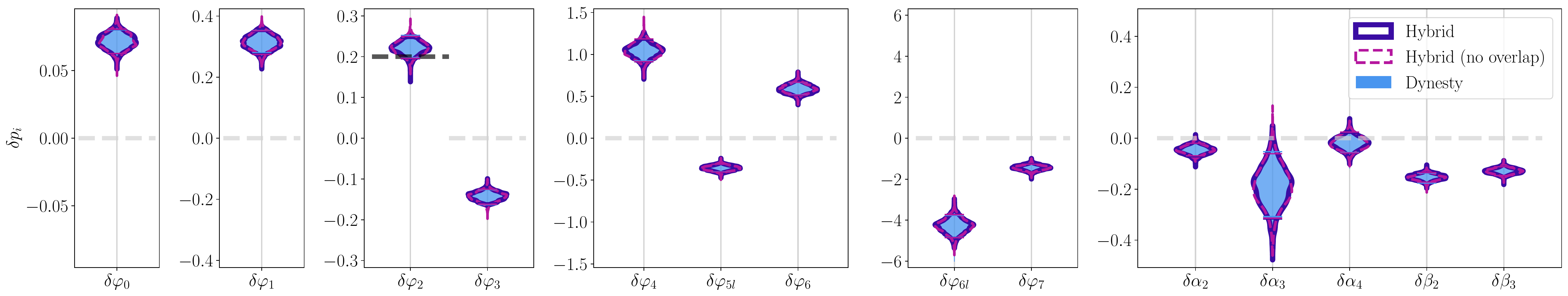}
    \caption{
        Violinplot, showing the posterior distributions on each post-Newtonian deviation parameter $\delta p_i$ for our injected signal, generated by hybrid sampling in the ``no overlap cut" and ``overlap cut" cases, with an additional solid-color posterior generated by using \code{dynesty} alone to check our results.
        Colored horizontal bars denote the 5th and 95th quantiles of the posteriors.
        In light gray, the injected value of $\delta p_i = 0$ is noted on each posterior with a dashed line, with a dark gray dashed line denoting the injected value of $\delta \varphi_2 = 0.2$.
        We observe that our hybrid sampling analysis agrees with \code{dynesty}-only analyses in all cases.
        Further, we observe that posteriors for $\delta \varphi_2$ generated by both methods are consistent with the injected value of $0.2$, but those posteriors for other parameters are incorrectly inconsistent with $0$.}
    \label{fig:highsnr_violinplot}
\end{figure*}

\begin{figure*}
    \centering
    \includegraphics[width=\linewidth]{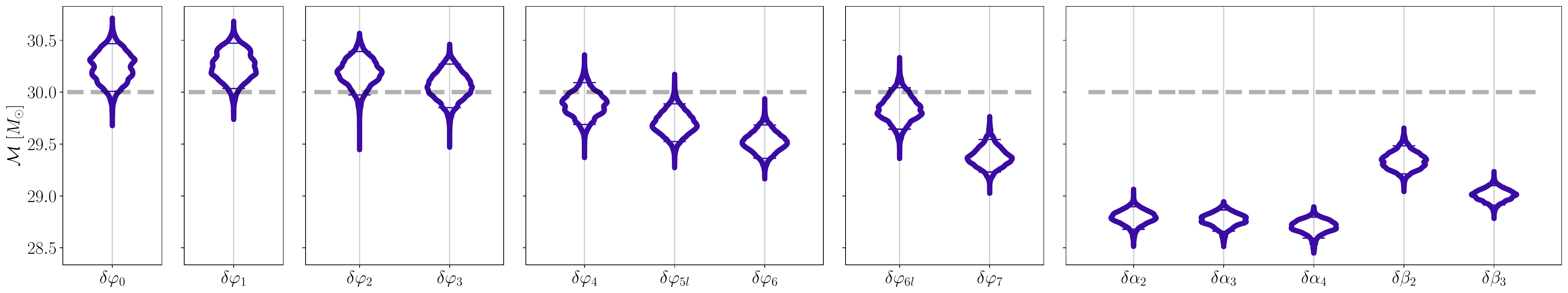}
    \caption{
        Violinplot, showing the posterior distributions on chirp mass generated while estimating each post-Newtonian deviation parameter $\delta p_i$ for our injected signal, generated by hybrid sampling in the ``overlap cut" case. Colored horizontal bars denote the 5th and 95th quantiles of the posteriors.
        In dark gray, the injected value of $\mathcal{M} = 30 M_{\odot}$ is noted on each posterior with a dashed line.
        We observe that our posterior distributions are \textit{only} consistent with the injected value when varying inspiral phase coefficients $\varphi_i$, which follows from their definition as degenerate with the mass of the system. This indicates that, if we were to receive a signal whose phase evolution disagreed with that predicted by general relativity, we would require a waveform model that admits deviations in $\delta p_i$ degenerate with the mass of the system.}
    \label{fig:chirp-mass_violinplot}
\end{figure*}

In Figure~\ref{fig:highsnr_corner} we show the one- and two-dimensional marginal posterior probability distributions for three parameters for this simulated signal.
From left to right (top to bottom) these are a non-GR deviation parameter $\delta \varphi_{2}$ and two intrinsic binary parameters, the chirp mass and mass ratio.
We note again that the deviation parameter is correlated with the intrinsic parameters.

In Figure~\ref{fig:highsnr_evolution}, we consider snapshots of the $\beta_{T}=1$ ensemble at various stages of the \code{ptemcee} analysis.
For this analysis, the injected chirp mass is strongly excluded from the posterior distribution obtained after the first, GR-only, analysis (pink) due to correlations between ${\cal M}$ and $\delta \varphi_{2}$.
As in Figure~\ref{fig:alpha-beta_evolution}, the ensemble of walkers evolves to explore the extended parameter space and converge on the correct solution.
As with our analysis of GW150914, $\sim 1000$ iterations are required until the ensemble converges.

In Figure \ref{fig:highsnr_violinplot}, we show the posteriors for $\delta p_i$ for our simulated signal.
Despite the deviation only being non-zero for $\delta \varphi_{2}$, the posterior distributions for all of the inspiral and intermediate deviation parameters are inconsistent with zero at high significance.
For the merger-phase deviation parameters, the deviations from zero are less pronounced.
This is consistent with previous work that has demonstrated that deviations at one post-Newtonian order can be identified with other deviation parameters~\cite{meidam2018, pi_haster2020} due to correlations between the parameters~\cite{Saleem2022}.

Additionally, in Figure \ref{fig:chirp-mass_violinplot}, we note that the posterior distributions for chirp mass we obtain while estimating $\delta p_i$ are \textit{only} consistent with the injected value when allowing one of the inspiral phase deviation coefficients $\delta \varphi_i$ to vary. In general, corrections at similar post-Newtonian orders are more strongly correlated, and this is visible from our results. Thus, if we receive a signal whose phase evolution is inconsistent with general relativity we cannot trust our estimate of the chirp mass and require a model with additional degrees of freedom to capture the mass term accurately.

Comparing the number of likelihood evaluations for each analysis, we find that each \code{dynesty} analysis requires $\sim 10^7$ likelihood evaluations and each \code{ptemcee} analysis requires $\sim 3 \times 10^6$ likelihood evaluations.
For each \code{dynesty} analysis we use only 500 live points in this case, compared to 2000 for our analysis of GW150914 and so we expect the \code{dynesty} analysis to require a factor of four fewer likelihood evaluations.
Taking this into account, we see a comparable (or even larger) computational saving with our hybrid method as for our analysis of GW150914.

\section{Conclusions} \label{sec:conclusions}
In this work, we introduced a novel hybrid sampling method for exploring models that can be described as extensions of a simpler underlying model.
By seeding a parallel-tempered ensemble MCMC with initial posterior estimates generated by performing nested sampling on a base model, hybrid sampling efficiently explores the extended parameter space of a more complex model.
While previous methods have employed similar hybrid sampling methods, e.g.,~\cite{Miller2019,Psaltis2021}, we exploit the athermal property of the nested sampling algorithm to optimally seed the ensembles of walkers at each temperature.

First, we demonstrated our framework with a toy model, using hybrid sampling to estimate the parameters of a generalized Gaussian distribution.
We saw that we are able to successfully recover the parameters of the true model, even when the base model is misspecified and the parameters of the extended model are correlated with those of the base model.

Following this, we applied our method to a widely performed test of general relativity with gravitational-wave transients: parameterized deviations from the waveform predicted by general relativity.
Using our method, we accurately reproduced the tests of general relativity using GW150914 as performed by the LIGO/Virgo scientific collaborations and estimate that our method is approximately an order of magnitude more efficient than the current direct sampling method \cite{single-BBH_gr-tests_LIGO_2016}.
Finally, we analyzed a simulated signal with a measurable deviation from the prediction of relativity.
We found that the efficiency of our hybrid sampling method is still far superior to direct sampling in this case.

Previous analyses have suffered from large computational costs as the parameters describing the waveform predicted by relativity are strongly correlated with the deviation parameters.
In order to mitigate this, we introduced a ``closeness'' criterion between the non-GR waveform being considered and the corresponding GR signal.
Specifically, this is implemented as a minimum overlap threshold between the two signals.
This acts as an additional prior constraint that the signal must be similar to the GR prediction, given the previous success of relativity.
This is particularly beneficial for lower signal-to-noise ratio systems where the data are less informative.

For the signals that we analyzed, we determined their consistency with GR by visual inspection of the marginal posteriors of each GR-deviation parameter returned by our hybrid analysis.
Although beyond the scope of this work, one quantitative test of the consistency of our results with GR is Bayesian model selection, wherein one would compare the evidences assuming no deviations from GR and allowing a deviation from GR.
The evidence from hybrid sampling is the evidence associated with the posterior generated by \code{ptemcee}; this code computes the evidence via thermodynamic integration \citep{goggans2004_thermo-integration, lartillot2006_thermo-integration} of the mean log-likelihood of each tempered chain \citep{Vousden_2015_ptemcee}.
In this work, we only used five temperatures, however, accurate calculation of the evidence would likely require more temperatures.
With accurate estimation of the evidences, one could compute a Bayes factor between the GR and beyond-GR waveform models.

While we have focused on a narrow application of measuring single additional parameters describing deviations from relativity, the method presented here can be used for more exploratory analyses that allow multiple non-GR parameters simultaneously that otherwise have exploding computational costs due to the number of possible combinations of parameters to vary simultaneously.
More generically, this method can be applied to any case where importance sampling to include a more physically realistic, but expensive model breaks down.
For example, measuring eccentricity in compact binary mergers~\cite{RomeroShaw2021}, estimating the impact of calibration uncertainty on inference~\cite{Payne2020}, and analyzing pairs of potentially gravitationally-lensed events~\cite{Janquart2022}.

\section{Acknowledgements}

    We thank Sylvia Biscoveanu, Max Isi, Nathan Johnson-McDaniel, Ralph Smith, Salvatore Vitale, and Alan Weinstein for helpful discussions and comments.
    CT is supported by an MKI Kavli Fellowship. JG is supported by PHY-1764464. 
    NW acknowledges support from the National Science Foundation (NSF) and the Park Scholarships program at NC State.
    We are grateful to the LIGO Caltech SURF program where this project began which is supported by the NSF REU program. 
    This material is based upon work supported by NSF's LIGO Laboratory which is a major facility fully funded by the National Science Foundation. The authors are grateful for computing resources provided by the California Institute of Technology and supported by National Science Foundation Grants PHY-0757058 and PHY-0823459.
    The analysis in this work made use of data available from the Gravitational Wave Open Science Center (gw-openscience.org)~\cite{gwosc}.
    This analysis used the following software: \code{numpy} \cite{numpy}, \code{scipy} \cite{scipy}, \code{matplotlib} \cite{matplotlib}, \code{corner.py} \cite{corner}, \code{pandas} \cite{reback2020pandas, mckinney-proc-scipy-2010}, \code{LALSimulation} \cite{lalsuite}, \code{bilby} \cite{Ashton2019, RomeroShaw2020}, \code{dynesty} \cite{speagle_2020_dynesty}, \code{ptemcee} \cite{Vousden_2015_ptemcee}.
    We provide analysis scripts, notebooks, and some data at \url{https://github.com/noahewolfe/tgr-hybrid-sampling}.
\bibliography{refs}{}

\appendix
\section{Comparison of Initialization Methods} \label{app:150914-direct-ptemcee}

Hybrid sampling uses a posterior generated via nested sampling to initialize a set of tempered-ensembles of MCMC walkers, however, it is also possible to initialize MCMC walkers near the parameters that yield the maximum likelihood.
Here, we repeat the analysis of GW150914 allowing $\delta \varphi_2$ to vary with no overlap cut, using \code{ptemcee} initialized with two common methods of MCMC initialization.
In the ``Prior" method, we initialize our ensembles with random samples drawn from the prior distributions detailed in Tables~\ref{tbl:gw150914_gr_priors} and~\ref{tbl:gw150914_dpi_distributions}.
In the ``Maximum Likelihood" method, we initialize our ensembles near the maximum likelihood point, which we compute here as the peak of the likelihood function in our GR-only analysis of GW150914 using \code{dynesty}.
For the initial values of $\delta \varphi_2$ in this run, we sample from a narrow truncated Gaussian distribution centered on zero.
With both methods, we again employed $250$ walkers at $5$ temperatures, as in our hybrid analysis of GW150914.

\begin{figure}
    \centering
    \begin{tabular}{c}
        \includegraphics[width=\linewidth]{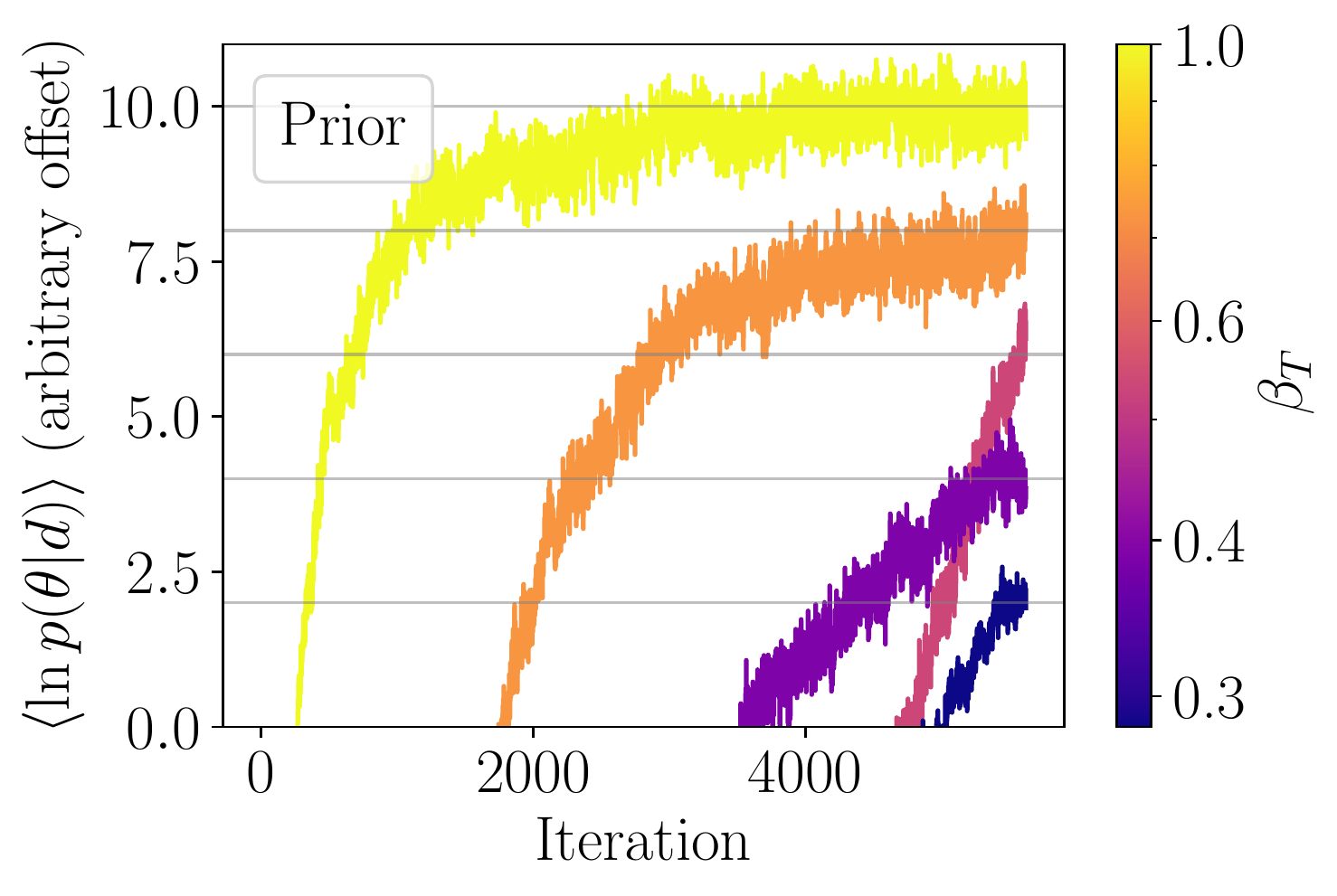} \\ \includegraphics[width=\linewidth]{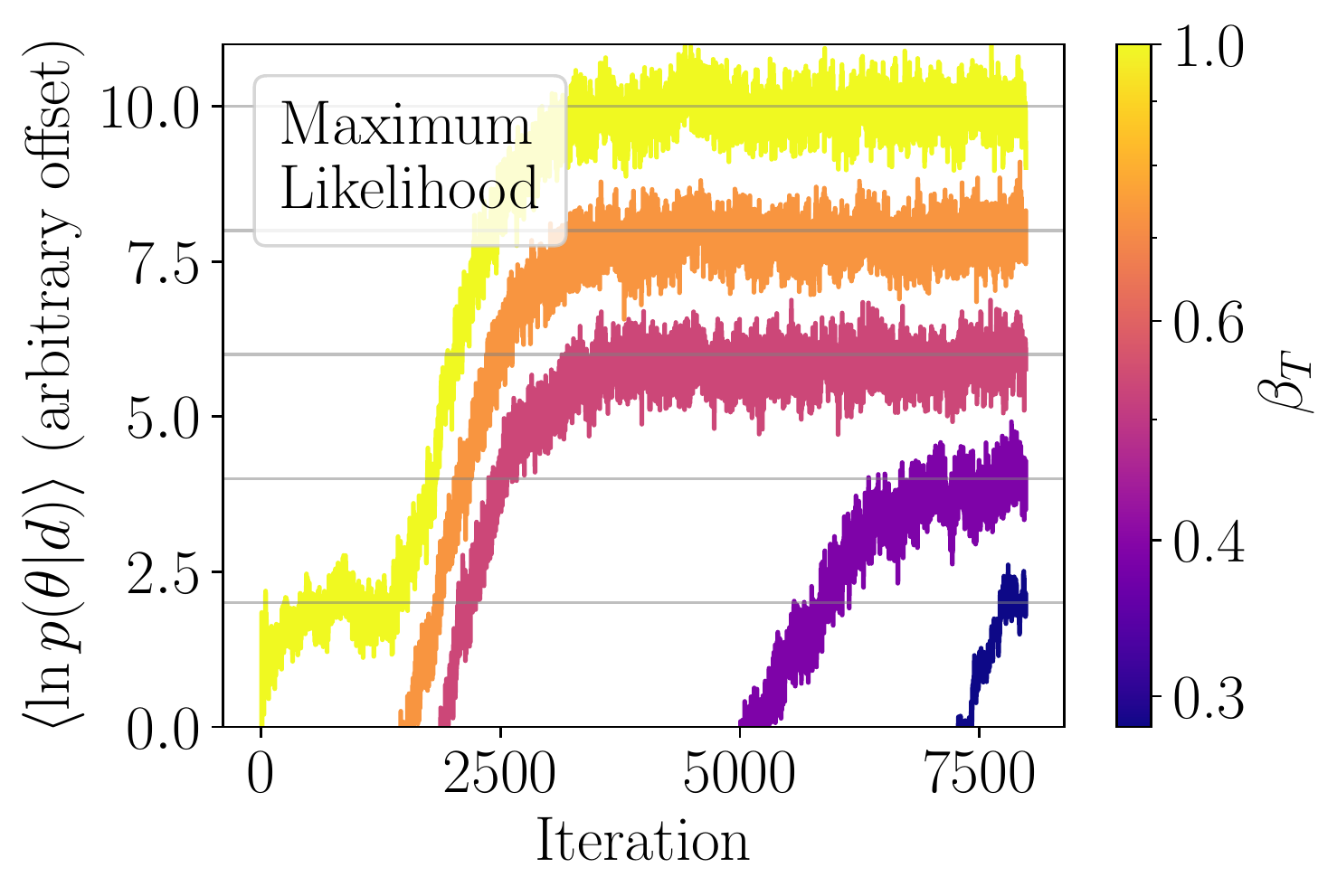} \\ \includegraphics[width=\linewidth]{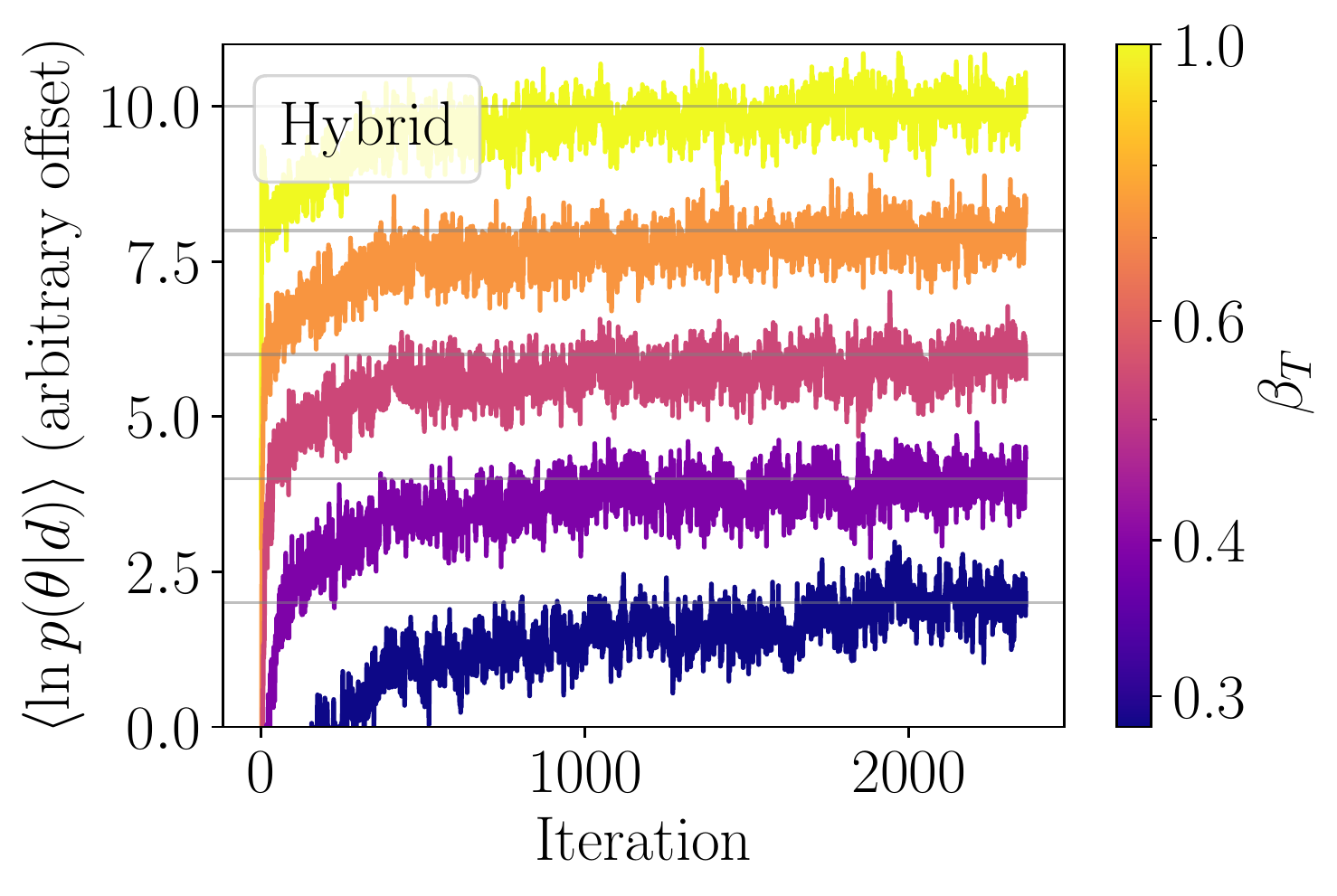}\\
    \end{tabular}
    
    \caption{The logarithm of the posterior probability averaged over all walkers at each temperature, $\langle \ln{p(\theta|d)} \rangle$, offset by an arbitrary value to simultaneously show $\langle \ln{p(\theta|d)} \rangle$ at each temperature. From left to right, we show $\langle \ln{p(\theta|d)} \rangle$ for the tempered-ensembles initialized with random samples from the prior, near the maximum likelihood point, and with our hybrid initialization method. We observe that only the hybrid initialized ensembles achieve convergence across all temperatures.}
    \label{fig:gw150914_init_comparison_meanlogposterior}
\end{figure}

So that the lowest-temperature $\beta_T = 1$ ensemble may more efficiently explore the entire target distribution, \code{ptemcee} proposes swaps between ensembles of different temperatures throughout their evolution.
While technically unlikely, this means in principle that swaps can occur between relatively hot and cold ensembles.
Therefore, to formally consider a set of tempered ensembles converged, the ensemble at each temperature must be converged, or else swaps between ensembles of different temperatures do not satisfy detailed balance (see \cite{Vousden_2015_ptemcee} and references therein for additional discussion).
In Figure~\ref{fig:gw150914_init_comparison_meanlogposterior}, we show the mean logarithm of the posterior probability density, $\langle \ln{p(\theta|d)} \rangle$, at each MCMC iteration and each temperature.
Generally, if $\langle \ln{p(\theta|d)} \rangle$ for a particular ensemble appears to be in a steady state, we expect that ensemble to have converged.
With this perspective in mind, Figure~\ref{fig:gw150914_init_comparison_meanlogposterior} indicates that the high-temperature (low $\beta_T$) ensembles initialized from the prior distribution or near the maximum likelihood point have failed to converge in over 4000 iterations.
In half that time, the ensembles initialized with our hybrid methodology have converged at every temperature.

\begin{figure}
    \centering
    \includegraphics[width=\linewidth]{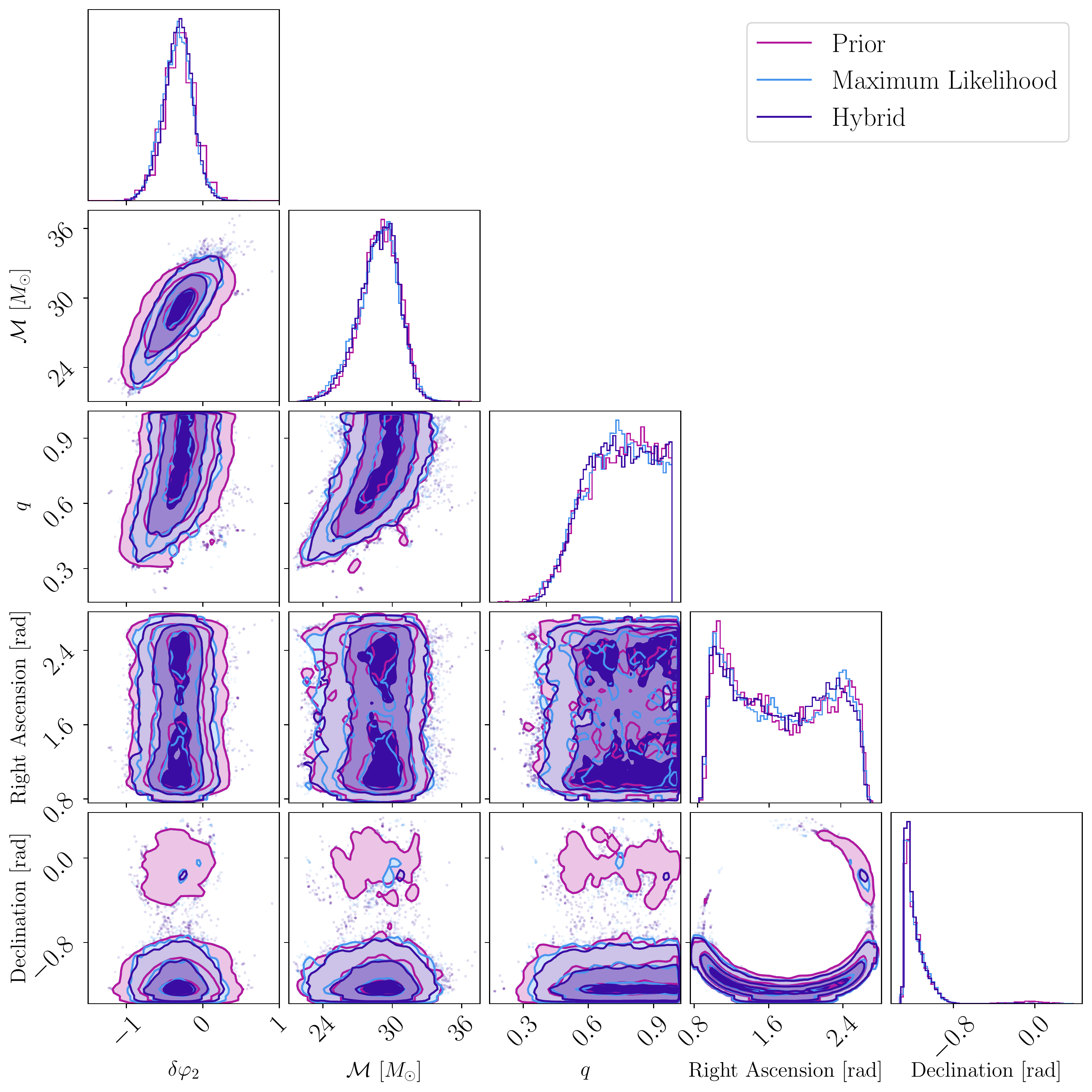}
    \caption{The posterior distributions generated via $\code{ptemcee}$ initialized with random samples from the prior (pink), near the maximum likelihood point (blue), and with our hybrid initialization method (purple). We observe that the results obtained with each initialization method are consistent with one another, however, the ``Prior" posterior distribution does not appear converged, consistent with the leftmost panel of Figure~\ref{fig:gw150914_init_comparison_meanlogposterior}.}
    \label{fig:gw150914_init_comparison_corner}
\end{figure}

The convergence of these sets of tempered ensembles has direct consequences for the posterior distributions that they yield.
In Figure~\ref{fig:gw150914_init_comparison_corner}, we compare the posterior distributions generated with each method for initializing \code{ptemcee}.
While the results obtained with each method are generally consistent, the ensembles initialized from the prior (pink) appear not converged with respect to the hybrid initialization (purple), particularly when looking at the marginal posteriors on declination.
This particular result can be explained by the uniform priors adopted for the sky position parameters, as the MCMC ensembles initialized from the prior generally had a much larger distance to travel across the likelihood surface compared to the ensembles initialized with our hybrid method, or even near the maximum likelihood point.
In total, we can conclude that our method yields results consistent with these standard methods of MCMC initialization while achieving convergence of the entire set of tempered MCMC ensembles in less than half the number of iterations.

\section{Sampler Settings} \label{sec:sampler_settings}

\begin{center}
\begin{table}
\begin{tabular}{|c c|}
 \hline
 Sampling Argument & Value \\ 
 \hline
 \multicolumn{2}{|c|}{\code{dynesty}} \\
 \hline
 \code{nlive} & $500$ \\
 \code{sample} & \code{`rwalk'} \\
 \code{walks} & 50 \\
 \code{nact} & 10 \\
 \hline
 \multicolumn{2}{|c|}{\code{ptemcee}} \\
 \hline
 \code{ntemps} & 5 \\
 \code{nwalkers} & 250 \\
 \code{burn\_in\_fixed\_discard} & 2000 \\
 \hline
\end{tabular}
\caption{Sampler arguments for hybrid sampling used in our analysis of injected signals as defined for the \code{Bilby} implementations of \code{dynesty} and \code{ptemcee}. For the \code{dynesty}-only analyses of injected signals, we also use \code{dynesty} sampler settings in this table.} \label{tbl:injected_sampler_kwargs}
\end{table}
\end{center}

To enable reproducibility of our results, we provide the settings used during each stage of our hybrid sampling algorithm.
These are listed in Table~\ref{tbl:injected_sampler_kwargs}.
All of the parameters are as defined in the \code{Bilby} implementation of the respective sampling code.
Additionally, configuration files can be found at \url{https://github.com/noahewolfe/tgr-hybrid-sampling}.
We note that for our analysis of GW150914, we used \code{nlive}$ = 2000$ rather than 500 as for the simulated signals.

\section{Further results for GW150914 Analysis}\label{app:150914-traces}

\begin{figure*}
    \centering
    \includegraphics[width=0.49\linewidth]{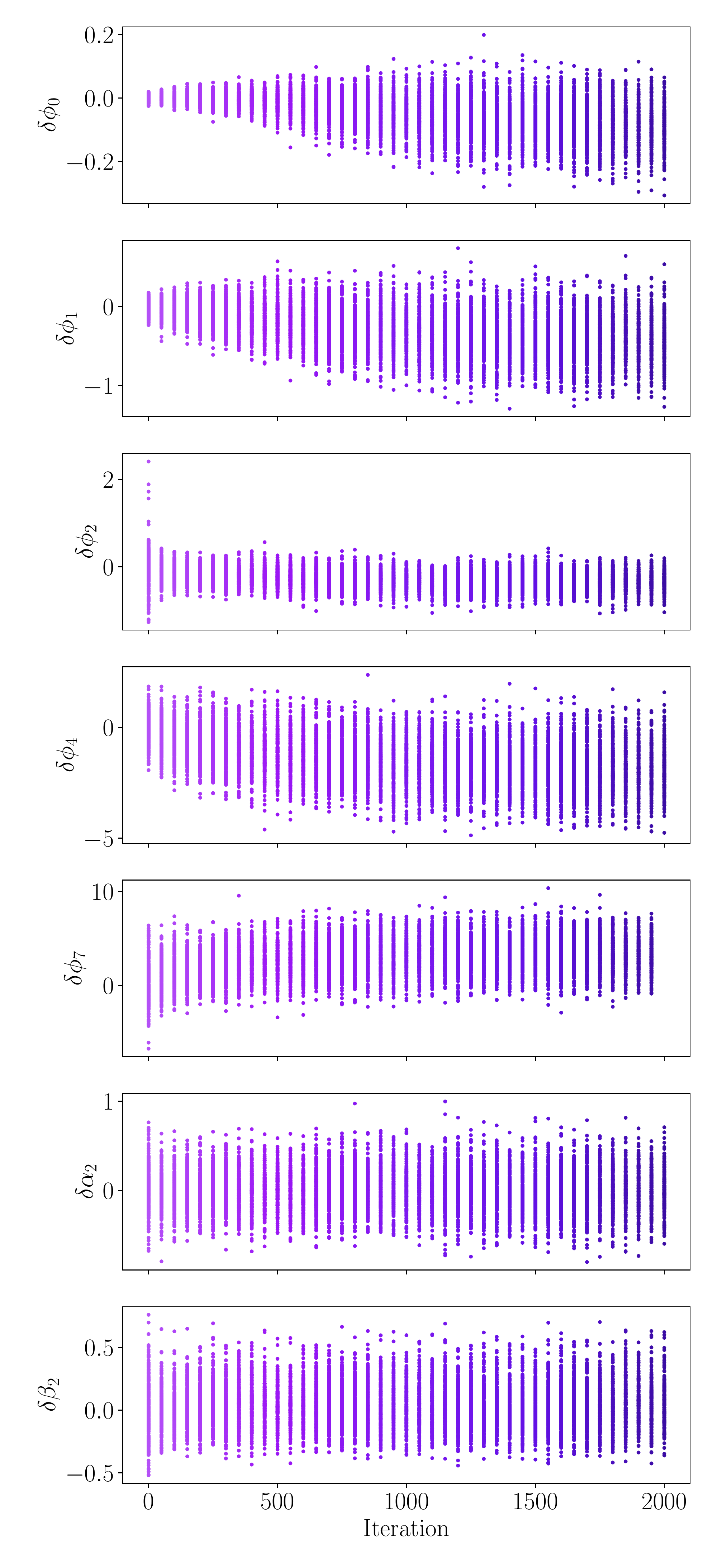}
    \includegraphics[width=0.49\linewidth]{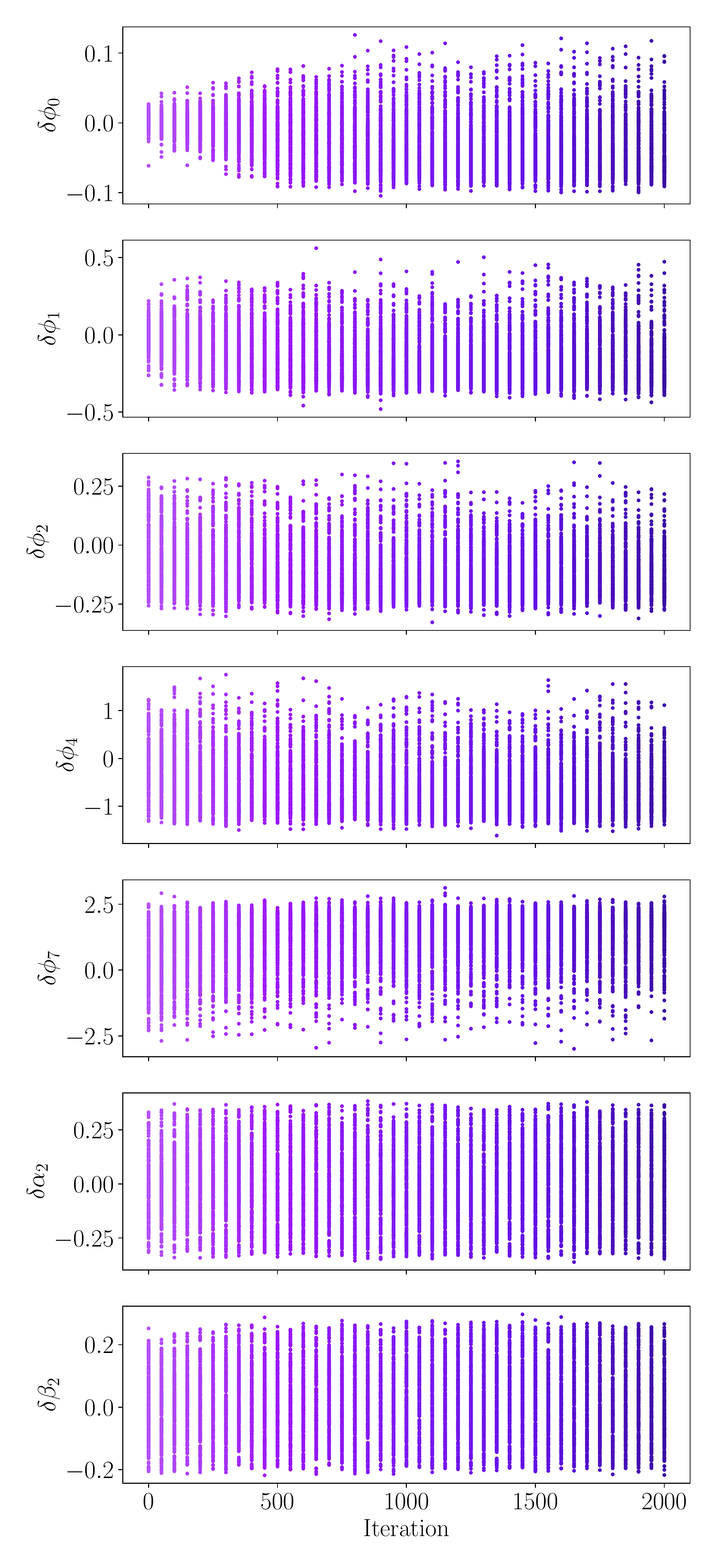}
    \caption{
    Trace plots showing the evolution of samples taken in a subset of post-Newtonian deviation coefficients for GW150914 from both the inspiral ($\delta \varphi_0$, $\delta \varphi_1$, $\delta \varphi_2$, $\delta \varphi_4$, $\delta \varphi_7$) and post-inspiral ($\delta \alpha_2$, $\delta \beta_2$) during the first 2000 steps of the second step of hybrid sampling. 
    Traces in the left column are generated without any overlap cut, whereas traces in the right column are generated while $\mathcal{O} > 0.9$ is imposed.
    The samples plotted at each iteration of sampling are collated from each of the 250 walkers in the ensemble at temperature $\beta_T = 1$.
    The color scheme matches the state of the ensemble shown in Figure~\ref{fig:gw150914_evolution_no-overlap}.
    We observe that even in the most extreme case, when no overlap cut is applied, all ensembles converge within $\sim1000$ iterations, with many converging far sooner particularly when an overlap cut is applied.
    }
    \label{fig:gw150914_trace}
\end{figure*}

In this appendix, we provide trace plots showing the evolution of the $\beta_{T}=1$ ensemble for the deviation parameters $\delta p$ for our analysis of GW150914.
In Figure~\ref{fig:gw150914_trace}, we show the results of analyses without (left) and with (right) the requirement that ${\cal O} \geq 0.9$ respectively.
In general, the sampler has converged to a steady-state after $\sim1000$ iterations and always after 2000 iterations.
We note that in most cases implementing our overlap condition reduces the number of iterations required for the ensemble to converge to a steady-state.

\section{Effect of Time-Maximized Overlap Cut} \label{app:time-maximization}

\begin{figure}[H]
    \centering
    \includegraphics[width=\linewidth]{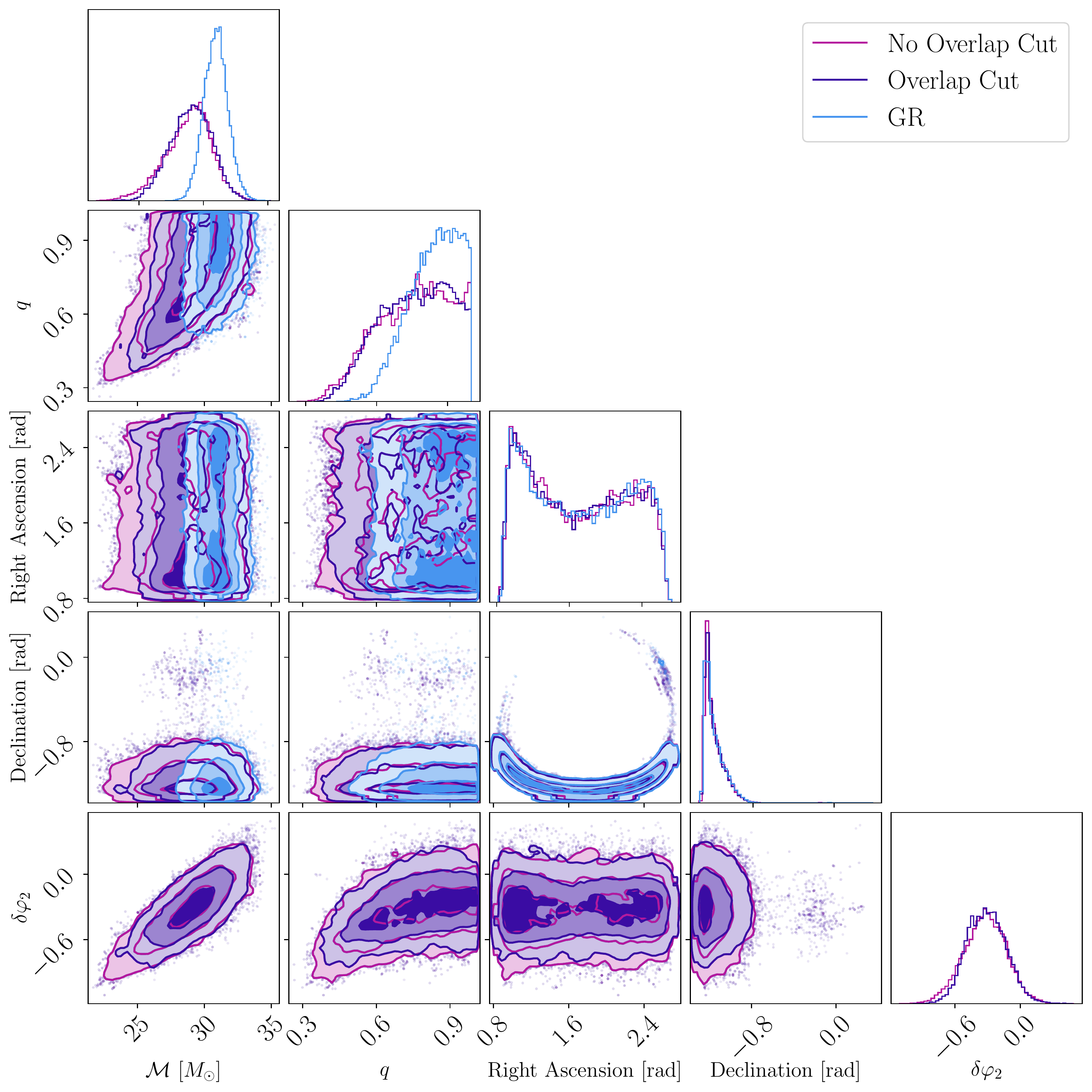}
    \caption{Joint posterior distributions on chirp mass $\mathcal{M}$, mass ratio $q$, right ascension, declination, and deviation $\delta \varphi_2$ during our estimation of $\delta \varphi_2$ in GW150914 with hybrid sampling. In magenta, we plot the posteriors with no overlap cut enforced, whereas in purple we enforce an overlap cut of $\mathcal{O} > 0.9$ which has been maximized over both merger phase and time, in contrast to the results presented in Figure~\ref{fig:gw150914_dphi2_overlap-comparison}. In blue, we show posteriors for $\mathcal{M}$ and $q$ from the first step of hybrid sampling, with no deviations from general relativity. In both hybrid results, we see the expected correlation between $\delta \varphi_2$ and $\mathcal{M}$, and lack of correlation between $\delta \varphi_2$ and the extrinsic sky parameters. Although the hybrid result with an overlap cut is constrained relative to its counterpart without that cut, the difference between these posterior distributions is visually smaller than when $\mathcal{O}$ was only maximized over the merger phase.}
    \label{fig:gw150914_dphi2_overlap-comparison_time-maximized}
\end{figure}

In Section~\ref{sec:bgr-waveforms}, we introduced the overlap $\mathcal{O}$ to measure the deviation in a waveform induced by a beyond-GR deviation, maximized over the merger phase of the signal. 
Changing the merger time $t_c$ introduces a frequency-dependent shift in the phase of the signal that is degenerate with a beyond-GR deviation; thus, some parametric tests of general relativity maximize $t_c$ as well when calculating the overlap (see, for example, \cite{ppE_bonilla_2022}).
In Figure~\ref{fig:gw150914_dphi2_overlap-comparison_time-maximized}, we present posterior distributions on the chirp mass, mass ratio, and the inspiral deviation parameter $\delta \varphi_2$ during our estimation of $\delta \varphi_2$ in GW150914, similar to the results presented in Figure~\ref{fig:gw150914_dphi2_overlap-comparison}.
Here, however, we have maximized $\mathcal{O}$ over both the merger phase and time.
This reduces the cut on the prior for $\delta \varphi_2$ imposed by requiring $\mathcal{O} > 0.9$, as a larger range of deviations in the waveform induced by $\delta \varphi_2$ can be accounted for by varying the merger time.
In turn, time maximization allows the mass parameters, in particular the chirp mass, to vary more widely as well, reducing the efficacy of an overlap cut.

\end{document}